\documentclass[aap]{imsart}

\arxiv{math.PR/0000000}

\doi{10.1214/154957804100000000}
\pubyear{0000}
\volume{0}
\firstpage{0}
\lastpage{0}

\RequirePackage[OT1]{fontenc}

\RequirePackage{amsthm,amsmath}
\RequirePackage[numbers]{natbib}
\RequirePackage[colorlinks,citecolor=blue,urlcolor=blue]{hyperref}

\usepackage{amsfonts}
\usepackage{algorithmic}

\usepackage{epsfig,array,bbm}
\usepackage[usenames,dvipsnames]{xcolor}

\newcommand{\vectornorm}[1]{\left\|#1\right\|}

\startlocaldefs

\def\defeq{\stackrel{\Delta}{=}}

\def\eps{\varepsilon}

\newtheorem{defn}{Definition}[section]

\newtheorem{remark}{Remark}[section]

\newtheorem{prop}{Proposition}[section]

\def\greg#1{\textcolor{ForestGreen}{#1}}

\newcommand{\R}{\mathbb R}

\def\NN{{\mathcal N}}

\def\beq{\begin{equation}}
\def\eeq{\end{equation}}
\def\bqn{\begin{eqnarray}}
\def\eqn{\end{eqnarray}}
\def\bqns{\begin{eqnarray*}}
\def\eqns{\end{eqnarray*}}
\def\bep{\begin{proof}}
\def\ep{\end{proof}}

\def\bc{\begin{center}}
\def\ec{\end{center}}

\def\s{\sigma}
\newcommand{\N}{\mathbb N}

\DeclareMathOperator*{\Hess}{Hess}
\DeclareMathOperator*{\diag}{diag}

\endlocaldefs

\begin{document}
\setattribute{journal}{name}{}

\begin{frontmatter}

\title{An adaptive ridge procedure\\ for $L_0$ regularization}
\runtitle{Adaptive Ridge regression }


\author{\fnms{Florian} \snm{Frommlet}\ead[label=e1]{Florian.Frommlet@meduniwien.ac.at}}
\address{
Department of Medical Statistics (CEMSIIS)\\
 Medical University of Vienna\\
  Spitalgasse 23 \\
A-1090 Vienna, Austria\\\printead{e1}}
\and
\author{\fnms{Gr\'egory} \snm{Nuel}\ead[label=e2]{nuel@math.cnrs.fr}}

\address{
National Institute for Mathematical Sciences (INSMI), CNRS\\
Stochastics and Biology Group (PSB), LPMA UMR CNRS 7599 \\
Universit\'e Pierre et Marie Curie\\
4 place Jussieu, 75005 Paris, France\\
\printead{e2}}

\runauthor{Frommlet and Nuel}

\begin{abstract}
Penalized selection criteria like AIC or BIC are among the most popular methods for variable selection. Their theoretical properties have been studied intensively  and are well understood, but  making use of them in case of high-dimensional data is difficult due to the non-convex optimization problem induced by $L_0$ penalties. An elegant solution to this problem is provided by the  multi-step adaptive lasso, where iteratively weighted lasso problems are solved, whose weights are updated in such a way that the procedure converges towards selection with $L_0$ penalties. In this paper we introduce an adaptive ridge procedure (AR) which mimics the adaptive lasso, but is based on weighted Ridge problems. After introducing AR its theoretical properties are studied in the particular case of  orthogonal linear regression. For the non-orthogonal case extensive simulations are performed to assess the performance of AR. In case of Poisson regression and logistic regression it is illustrated how the iterative procedure of AR can be combined with iterative maximization procedures. The paper ends with an efficient implementation of AR in the context of least-squares segmentation.

\end{abstract}





\end{frontmatter}

\title{}



\maketitle

\begin{abstract}

\greg{todo}

\end{abstract}



\section{Introduction}\label{Sec:Intro}

Methods for performing variable selection, particularly in a high dimensional setting, have undergone tremendous development over the last two decades. Of particular importance in this context is penalized maximum likelihood estimation, which can be divided in selection methods based on generalized information criteria and regularization methods \cite{Chen2014}. The former use a penalty which depends on the number of estimated parameters, sometimes called $L_0$ penalty, and include the classical information criteria  AIC \cite{A74} and BIC \cite{S78}. Their asymptotic properties have been thoroughly studied and are well understood when the number of potential regressors is fixed (see for example \cite{Y05} and citations given there). Specifically BIC is known to yield a consistent model selection rule, which means that as the sample size goes to infinity the probability of selecting the true model goes to 1. However, this is no longer true in a high dimensional setting, where under sparsity both AIC and BIC tend to select too large models \cite{BS02}. As a consequence a number of different modifications of BIC have been suggested, for example mBIC \cite{BGD, BZG08} which is designed to control the family wise error rate (FWER), mBIC2 \cite{FBMC, FRTB12} controlling the false discovery rate, or EBIC \cite{Chen08} for which consistency under certain asymptotic conditions has been shown even when the number of regressors is allowed to be larger than the sample size.

Thus from a theoretical perspective it is rather appealing to perform model selection using generalized information criteria. However, the corresponding optimization problem is notoriously difficult due to the non-convexity and discontinuity of the $L_0$ penalty. It is an NP hard problem to find the model which minimizes a specific  information criterion, and in general already for a moderate number of say fifty variables it becomes computationally infeasible to guarantee finding the optimal solution. Another problem often associated with $L_0$ penalties is the instability of selected solutions \cite{B96}. A possible workaround is to report not only one model which minimizes the criterion, but a number of good models which have been found for example with some evolutionary algorithms \cite{FLAB}. In any case the approach remains extremely computer intensive and time consuming for high-dimensional data sets.

Regularization methods can serve as an alternative, where penalties are not based on the number, but rather on the size of coefficients. A prominent example is bridge regression \cite{frank1993statistical} which uses penalties of the form $\sum_i \beta_i^q$, where $\beta_i$ are the coefficients of the model to be estimated. Special cases are ridge regression \cite{HK70} for $q=2$  and the Lasso \citep{tibshirani1996regression} for $q = 1$, whereas for $q \rightarrow 0$ the penalty of bridge regression converges towards the $L_0$ penalty of generalized information criteria.  It has been shown that only for $q \leq 1$ bridge regression can perform variable selection \cite{KF00}, on the other hand only for $q \geq 1$ its penalty is convex and  therefore allows for relatively simple optimization algorithms. This partly explains the huge interest that the Lasso ($q=1$) has received in recent years (see \cite{BG11} for a comprehensive treatment).

The Lasso has very nice properties in terms of prediction, but as a model selection procedure it is consistent only under rather restrictive assumptions 
\citep{zhao2006model, BG11}. Specifically for strongly correlated regressors it can perform quite poorly, and a number of non-convex penalties have been studied to achieve sparser solutions \cite{MFH11}. Furthermore the coefficient estimates of the Lasso are severely biased due to shrinkage. An interesting procedure to overcome these deficits is the adaptive Lasso  \citep{zou2006adaptive}, which makes use of a weighted $L_1$ norm penalty resulting in a similar convex optimization problem as the original Lasso. With suitable choice of the weights the adaptive Lasso was shown to have the oracle property, which means that it is both consistent and the nonzero coefficients are estimated as well as when the correct model was known. The weights for the adaptive Lasso can be obtained with some initial Lasso estimates, and if this procedure is further iterated one obtains a multi-step adaptive Lasso \cite{BM08,candes2008enhancing}

Already much earlier Grandvalet showed that the Lasso estimate can be obtained via some weighted ridge regression  \citep{grandvalet1998least,canu1999outcomes}. He called his procedure adaptive ridge regression, of which a slightly modified version has been recently applied to detect rare variants in genome wide association studies \citep{zhan2012adaptive}.
In this article we want to study a different adaptive ridge procedure, which was recently proposed \citep{rippe2012vizualization} with the aim of approximating $L_0$ penalties. 
This Adaptive Ridge (AR) procedure is somewhat similar to the multi-step adaptive Lasso, in the sense that the weights are iteratively adapted; but in each iteration weighted ridge regression is performed instead of weighted Lasso, which is computationally much easier. 

The iteratively adapted weights of AR are designed in such a way that the resulting penalty converges towards the $L_0$ penalty. Therefore the procedure is somewhat related to the seamless $L_0$-penalty \cite{Dicker2013}    
and the combination of $L_0$ and $L_1$ penalties suggested in \cite{LW07}, which both represent regularized versions of the $L_0$ penalty. However, the latter procedures rely upon non-convex optimization, which gets computationally rather difficult for large-scale problems as well as for applications beyond linear regression. In contrast each iteration of the suggested AR is extremely fast, and we will see that the method also performs really well in some non-linear examples.

The main purpose of this article is to look more systematically into the statistical properties of the AR procedure proposed in \citep{rippe2012vizualization}. After introducing the general procedure in Section \ref{Sec:GenProc}, we will focus in Section \ref{Sec:Linear} on the special case of linear regression. In particular we will provide some theoretical results on the behavior of AR under an orthogonal design, and we will show to which extent these results apply for more general design matrices. In Section \ref{Sec:Ex} the performance of AR will be studied for generalized linear models and for least squares segmentation. We finally end with a discussion in Section \ref{Sec:Disc}.

\section{General Procedure}\label{Sec:GenProc}

\subsubsection*{The Problem}

Consider a parametric model with parameter vector $\boldsymbol{\beta} \in \mathbb{R}^{d}$, in combination with a $\mathcal{C}^2$ convex contrast $C:\mathbb{R}^ d \rightarrow \mathbb{R}$. The most common examples of contrasts $C(\boldsymbol \beta)$ are the residual sum of squares, or minus twice the log-likelihood of a given model, but more general functions like pseudo-likelihood related quantities are conceivable.
  For all $0 \leq q \leq 2$, $\lambda \geq 0$ we introduce the penalized contrast
\begin{equation}\label{eq:pencontrast}
C_{\lambda,q}(\boldsymbol \beta) \defeq 
C(\boldsymbol \beta)+{\lambda} \vectornorm{ \boldsymbol \beta}_{L_q}^q \; .
\end{equation}
\begin{remark} \label{Rem:mat}
One can easily replace $\boldsymbol \beta$ in the penalty term by any linear transformation $\boldsymbol {D \beta}$ allowing to consider wider generalizations of penalty forms. For example one might consider a subspace extraction such that  $\boldsymbol D\boldsymbol \beta =\boldsymbol \beta_\mathcal{J}$ for a given set $\mathcal{J}\subset\{1,2,\ldots,d\}$, or a difference matrix such that $\boldsymbol D \boldsymbol \beta=(\beta_1-\beta_2, \beta_2-\beta_3, \ldots, \beta_{d-1}-\beta_{d})^T$ (where~$^T$ denotes the transpose operator). We will make use of this only in the example of Section \ref{Subsec:Seg}. All the results obtained previously can be straightforwardly extended for penalties of the form $\vectornorm{ \boldsymbol {D \beta } }_{Lq}^q$, but the generalization is omitted for the sake of simplicity.
\end{remark}

The objective of this paper is to minimize the penalized contrast of equation~(\ref{eq:pencontrast}) in order to obtain:
\begin{equation}\label{eq:betapen}
\hat{\boldsymbol \beta}
\defeq \arg \min_{\boldsymbol \beta} C_{\lambda,q}(\boldsymbol \beta).
\end{equation}
This relates to Bridge regression for $q > 0$ \citep[see][]{frank1993statistical}, with the special cases of ridge regression for $q=2$, and LASSO for $q =1$ \cite{tibshirani1996regression}. Note that if $q>1$, the penalized contrast is both convex and $\mathcal{C}^2$ and the problem can be easily solved with straightforward  convex optimization (Gradient descent, Newton-Raphson, etc.). For $q=1$, the problem is still convex but with derivative singularities that makes the optimization problem more delicate but still tractable (coordinate descent \cite{wu2008coordinate}, gradient LASSO \cite{kim2008gradient}, etc.). If $0 \leq q <1$, the penalized contrast is not convex anymore and the problem is much more challenging \cite{MFH11}. For the limiting case $q = 0$ one obtains for suitable choices of $\lambda$ the classical model selection criteria AIC and BIC. Only for very small $p$ it is possible to apply exact algorithms which guarantee to find the minimal solution \citep{furnival1974regressions}, whereas for $p > 20$ one essentially has to use heuristic search strategies like stepwise selection procedures. However, variable selection based on $L_0$ penalties is believed to be optimal for achieving sparsity and unbiasedness, and therefore there is much interest to find efficient algorithms which minimize $C_{\lambda,q}$ also in case of $q = 0$.

\subsubsection*{The Suggested Solution}

Recently Rippe et al. \cite{rippe2012vizualization} suggested  a method for visualizing changes of copy number variation along the chromosome which is based on an iterative procedure to minimize residual sum of squares with $L_0$ penalties. We will adapt this procedure to our setting of penalized likelihoods and discuss it in a slightly more general form. The idea is to obtain $\hat{\boldsymbol\beta}$ 
through an iterative weighted fixed-point procedure. For any $\lambda \geq 0$  and any non-negative weight vector $\boldsymbol w \in \mathbb{R}_+^p$ we introduce the function:
\begin{equation}\label{eq:penARcontrast}
F_{\lambda,\boldsymbol w}(\boldsymbol\beta) \defeq C(\boldsymbol\beta)
+\frac{\lambda}{2} \boldsymbol\beta ^T \diag(\boldsymbol w) \boldsymbol \beta
=C(\boldsymbol\beta) + \frac{\lambda}{2} \sum_{j=1}^d w_j \beta_j^2 \; ,
\end{equation}
where $\diag(\boldsymbol w)$ is the diagonal matrix with weights $\boldsymbol{w}$ on its diagonal. We are now ready to introduce our Adaptive Ridge procedure:
\begin{defn}[AR]\label{def:AR}
For any  $\lambda>0$ and $0 \leq q < 2$, the $L_q$ Adaptive Ridge sequences $\boldsymbol \beta ^{(k)}$ and $\boldsymbol w ^{(k)}$  are defined by the initialization $\boldsymbol w^{(0)} = \boldsymbol 1$, and for $k \in \N$ by:
\begin{equation}\label{eq:argmin}
\boldsymbol \beta ^{(k)} = \arg \min_{\boldsymbol \beta} F_{\lambda,\boldsymbol w^{(k-1)}}(\boldsymbol\beta)
\end{equation}
\begin{equation}\label{eq:weights}
\boldsymbol w^{(k)} = \left( \left| \boldsymbol \beta ^{(k)} \right|^\gamma+\delta^\gamma
\right)^{(q-2)/\gamma}
\end{equation}
where Equation~(\ref{eq:weights}) is defined component-wise, and depends on the constants $\delta > 0$ and $\gamma > 0$.
\end{defn}
Equation~(\ref{eq:argmin}) is just a weighted version of ridge regression, which is usually fast to solve. Note that for $q = 2$ one always has $\boldsymbol w ^{(k)} = \boldsymbol 1$ and thus the procedure is not really iterative. In contrast for $q < 2$, $\boldsymbol w ^{(k)}$  \emph{does} depend on the iteration step $k$, and in case of convergence of the sequence $\boldsymbol \beta^{(k)}$ we will  write  $\boldsymbol \beta^{(k)} \rightarrow \tilde{\boldsymbol \beta}$.  

The form of the weights $\boldsymbol w^{(k)}$ of Equation~(\ref{eq:weights}) is motivated by the heuristic consideration that at least formally the penalty term of (\ref{eq:penARcontrast}) converges towards the penalty term of (\ref{eq:pencontrast}),
\beq \label{Eq:NormApprox}
\boldsymbol\beta ^{(k)T} \diag\left(\boldsymbol w^{(k-1)}\right) \boldsymbol \beta^{(k)}
\underset{k \rightarrow \infty}{\rightarrow}
\sum\limits_{j=1}^d \frac{\tilde \beta_j^2}{(|\tilde \beta_j|^\gamma+\delta^\gamma)^{\frac{(2-q)}{\gamma}} } \approx \sum\limits_{j=1}^d  |\tilde \beta_j|^q
= \vectornorm{ \tilde{ \boldsymbol \beta}}_{L_q}^q\; .
\eeq


For $q=1$ one obtains in the limit the Lasso penalty by iteratively solving weighted ridge problems, which has been exactly the motivation of the Adaptive Ridge Approach introduced in \cite{grandvalet1998least}.  However, the main aim of our Adaptive Ridge procedure AR is not to approximate the Lasso, but to focus on $0 \leq q < 1$, and especially on the case $q=0$. As a consequence our AR is more similar in spirit  to the multi-step adaptive Lasso discussed in \cite{BM08} and \cite{candes2008enhancing}, where iteratively the weights of the $\ell_1$ penalty are updated using formulas which are very similar to equation~(\ref{eq:weights}). More precisely both references make use of $\gamma=1$, whereas we will later recommend to work with $\gamma = 2$. Furthermore one finds $\delta=0$ in \cite{BM08}, whereas  \cite{candes2008enhancing} introduces $\delta>0$ for numerical stability.  Again we will discuss the exact choice of $\delta$ in our procedure below. 

The main advantage of our AR approach compared with the multi-step adaptive Lasso is that solving a ridge problem in each iteration is much easier than solving a lasso problem. While AR works for any $q < 1$ we will focus here on the case $q = 0$, which corresponds to a number of widely used variable selection criteria,  and for which minimizing (\ref{eq:pencontrast}) is particularly difficult. In fact this optimization problem is NP hard with growing $p$, and thus it is very useful to have a good approximate procedure.

\subsubsection*{Numerical considerations}

\begin{figure}[htb!]
\begin{center}
\includegraphics[width=\textwidth]{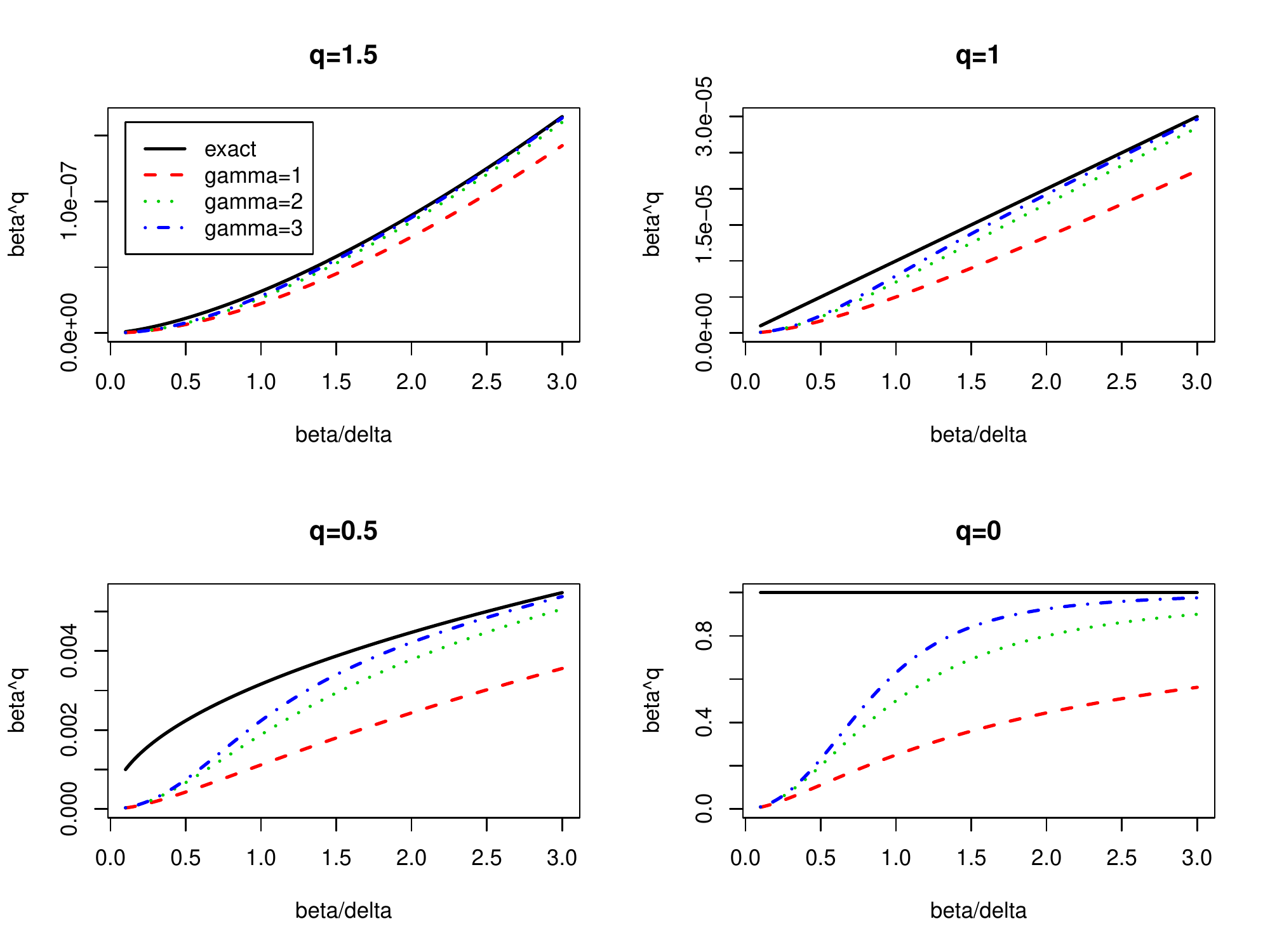}
\end{center}
\caption{Approximation of $|\beta_j|^q$   by the function  ${\beta_j^2}{(|\beta_j|^\gamma+\delta^\gamma)^{(q-2)/\gamma}}$ in dependence of the parameter $\gamma \in \{1,2,3\}$.  The x-axis is at the scale of $\delta$. The four  panels illustrate the cases  $q=1.5, 1, 0.5, 0$.
}
\label{fig:delta}
\end{figure}

In order to avoid numerical instabilities (mainly due to floating point arithmetics), we suggest to use  instead of (\ref{eq:weights}) the following formula to update the weights $w_j$ :
\beq \label{Eq:wFloatingPoint}
w_j = \left\{ 
\begin{array}{ll}
\delta^{q-2} \exp \left[ \frac{q-2}{\gamma}  \mbox{log1p}\left( \left|\frac{ \beta_j}{\delta}\right|^\gamma\right)\right]
& \text{if $|\tilde\beta_j| \leq \delta$} \\[2mm]
|\tilde\beta_j|^{q-2} \exp \left[ \frac{q-2}{\gamma}  \mbox{log1p}\left( \left|\frac{\delta}{\beta_j}\right|^\gamma\right)\right]
& \text{if $|\tilde\beta_j| > \delta$} \\
\end{array}
\right.
\eeq
where $\mbox{log1p}$ is the classical function defined by $\mbox{log1p}(u)\defeq \log(1+u)$ (for all $u>-1$).

According to Definition \ref{def:AR} the AR procedure depends  on two parameters, $\delta$ and $\gamma$. The choice of $\delta$ calibrates which effect sizes are considered as relevant. If $\beta_j < \delta$ the corresponding weight $w_j$ will become large. Eventually one will obtain in the limit $\tilde \beta_j \approx 0$, and thus also $w_j \tilde \beta_j^2 \approx 0$. On the other hand for  $\beta_j \gg \delta$ it holds that $w_j \tilde \beta_j^2 \approx |\tilde \beta_j|^q$. A choice of $\delta=0$  (like in \cite{BM08})  might then appear to be reasonable, but our numerical experiments show that it leads to numerical instabilities and that a small $\delta>0$ (like in \cite{candes2008enhancing,rippe2012vizualization}) performs noticeably better. Simulation results (not presented in this manuscript) suggest that the procedure is not particularly sensitive to the exact choice of $\delta$, which coincides with the findings of \cite{candes2008enhancing} in case of adaptive lasso.  Throughout this paper we will thus work with  $\delta = 10^{-5}$.

The second parameter  $\gamma$ determines the quality of the approximation $w_j \tilde \beta_j^2 \approx |\tilde \beta_j|^q$. Figure~\ref{fig:delta} illustrates for several choices of  $q$ the shape of $ w_j \tilde \beta_j^2$ depending on the parameter $\gamma$. Clearly for increasing values of $\gamma$ the approximation is getting closer to the desired thresholding step function. In simulations not presented here we observed dramatic improvement of the performance of AR by raising the parameter from $\gamma=1.0$ (like in \cite{BM08, candes2008enhancing, grandvalet1998least}) to $\gamma=2.0$ (like in \cite{rippe2012vizualization}), while further increasing of $\gamma$ did not yield much more benefit. 

For the rest of the paper we will focus on the variable selection case $q=0$, and stick with the choice $\delta=10^{-5}$ and $\gamma = 2$. The Adaptive Ridge Regression procedure for $L_0$ regularization is therefore defined by the following (component-wise defined) weighting scheme:
\begin{equation}\label{eq:weightsL0}
\boldsymbol w^{(k)} = \left( \left( \boldsymbol \beta ^{(k)} \right) ^2+\delta^2
\right)^{-1}.
\end{equation}

Finally it is interesting to point out that AR can easily cope with situations where solving the weighted ridge problem requires some iterative numerical algorithm for optimization, like gradient descent, Newton-Raphson, Marquardt, etc. The idea is simply to update the current value of $\boldsymbol \beta ^{(k)}$ through the iterative numeric procedure rather than computing the exact solution to the ridge problem in each step. In other words we propose to mix the iterative schemes of AR and the optimization algorithm. For example, the Newton-Raphson version of our procedure can be described as follows:
\begin{defn}[Newton-Raphson Adaptive Ridge]
For any $\lambda>0$ the Newton-Raphson Adaptive Ridge sequences $\boldsymbol \beta ^{(k)}$ and $\boldsymbol w ^{(k)}$  are defined by the initialization $\boldsymbol \beta^{(0)} = \boldsymbol 0$ and $\boldsymbol w^{(0)} = \boldsymbol 1$, and for $k \in \N$ by
\begin{equation} \label{eq:NRAR}
\boldsymbol \beta ^{(k)} =  \boldsymbol \beta ^{(k-1)} - \left[\Hess F_{\lambda,\boldsymbol w^{(k-1)}}\left(\boldsymbol\beta ^{(k-1)}\right) \right]^{-1}
\nabla F_{\lambda,\boldsymbol w^{(k-1)}}\left(\boldsymbol\beta ^{(k-1)}\right)
\end{equation}
with weights $\boldsymbol w^{(k)}$ being updated according to equation~(\ref{eq:weightsL0}).
\end{defn}


\section{Linear Regression} \label{Sec:Linear}

In this section we will systematically study AR with $q = 0$ as a variable selection procedure in the context of linear regression. 
Thus consider the model 
\beq \label{Eq:LinModel}
\boldsymbol y = \boldsymbol  X \boldsymbol  \beta + \boldsymbol  \eps \; ,
\eeq
where $\boldsymbol y \in \R^n$,   $\boldsymbol X = (\boldsymbol X_1,\dots,\boldsymbol X_p) \in \R^{n \times p}$ and $\boldsymbol \beta \in \R^p$. The error terms are assumed to be i.i.d. normal, $\eps_i \sim \NN(0, \s^2)$. Furthermore let $\boldsymbol y$ be centralized, that is $\sum_{i=1}^n y_i = 0$,  and let all regressors  be centralized and standardized such that $\boldsymbol X_j^T \boldsymbol X_j = n$. Specifically this means that we consider only models without intercept.  

 Clearly the log-likelihood of model (\ref{Eq:LinModel}) is of the form 
$$
\ell(\boldsymbol \beta, \s^2)  = \text{const.} - n \log \s - \frac 1{2\s^2} (\boldsymbol X \boldsymbol \beta -\boldsymbol  y)^T(\boldsymbol X\boldsymbol  \beta - \boldsymbol y) \; .
$$
Then $-2 \ell$ takes the role of the convex contrast $C$ in (\ref{eq:pencontrast}), and in case of known error variance $\s^2$ we obtain (after neglecting constants) 
$$
C(\boldsymbol \beta) = \frac 1 {\s^2} (\boldsymbol X \boldsymbol \beta - \boldsymbol y)^T(\boldsymbol X \boldsymbol \beta - \boldsymbol y) \defeq \frac{\mbox{RSS}(\boldsymbol \beta)}{\s^2} \;.
$$
Variable selection with classical model selection criteria like AIC or BIC becomes a special case of (\ref{eq:pencontrast}) with  $q = 0$. More specifically let a model be defined by the set of non-zero coefficients $M = \{j: \beta_j \neq 0\}$. Then (\ref{eq:pencontrast}) becomes
\beq \label{Eq:PenLogLik}
C_{\lambda,0}(\boldsymbol \beta)  = \frac 1{ \sigma^2} \mbox{RSS}(\boldsymbol \beta) + \lambda |M| \; ,
\eeq
which for a given model $M$ is clearly minimized at  $\hat {\boldsymbol \beta}_M$, the maximum likelihood estimate with respect to the given model.

We now want to compare variable selection based on (\ref{Eq:PenLogLik}) with the AR procedure defined by (\ref{eq:argmin}) and (\ref{eq:weightsL0}). It is straight forward to see that for linear regression (\ref{eq:argmin}) can be  written as an explicit dynamic system, 
\beq \label{Eq:ItProc3}
\tilde {\boldsymbol\beta}^{(k)} =  \left(\boldsymbol X^T \boldsymbol X + \tilde \lambda \s^2 \ \diag(\boldsymbol w) \right)^{-1} \boldsymbol X^T \boldsymbol y \; .
\eeq
One major result of this section is concerned with shrinkage of coefficients resulting from the AR procedure. It turns out that  the non-zero coefficients of $\tilde{\boldsymbol\beta} = \lim \tilde {\boldsymbol\beta}^{(k)}$ obtained via (\ref{Eq:ItProc3}) are smaller in absolute terms than the  maximum likelihood estimates $\hat {\boldsymbol\beta}^M$ of a model $M$ containing exactly the same non-zero coefficients as $\tilde{\boldsymbol\beta}$. Closely related is the fact that AR with parameter $\tilde \lambda = \lambda$ does not directly correspond to variable selection based on minimizing $C_{\lambda,0}(\boldsymbol\beta)$, but that a smaller value of $\tilde \lambda$ must be chosen.  We will first give a theoretical presentation of these results for orthogonal regressors, and then illustrate the situation for the general non-orthogonal case based on simulation results.

\subsection{Orthogonal case} \label{SubSec:Orthogonal}

Assume that $p \leq n$ and that the design matrix fulfills  $\boldsymbol X^T \boldsymbol X = n \boldsymbol I_p$, where $\boldsymbol I_p$ is the identity matrix of dimension $p$. Then the usual maximum likelihood estimate of $\beta$ for the saturated model becomes $\hat {\boldsymbol \beta} = \frac 1n \ \boldsymbol X^T \boldsymbol y$, and 
(\ref{Eq:PenLogLik}) evaluated at the maximum likelihood estimate of any given model can be rewritten as
$$
C_{\lambda,0}\left(\hat{\boldsymbol\beta}^M \right) =
\frac 1{ \sigma^2} \left (\boldsymbol y^T \boldsymbol y - n \sum\limits_{j\in M} \hat \beta_j^2 \right) + \lambda |M| \; .
$$
 Thus the penalized likelihood is minimized when all those regressors enter the model for which 
\beq \label{Eq:Thresh1}
\hat \beta_j^2 > \lambda \s^2  /n  \; ,
\eeq 
which results in the well known fact that under orthogonality the model selection approach defined by (\ref{Eq:PenLogLik}) is nothing else but a thresholding procedure for the individual coefficients.
Note that the whole argument relies upon the fact that in case of orthogonality the coefficients $\hat \beta_j$ are estimated independently from each other.

We next argue that AR also becomes a simple thresholding procedure under orthogonality. First note that
 (\ref{Eq:ItProc3}) can be rewritten as
\beq \label{Eq:ShrinkOrtho}
\tilde\beta_j^{(1)} = \frac{1}{1 + K} \hat \beta_j\;  
, \quad
\tilde\beta_j^{(k)} = \frac{1}{1 + \frac{K}{\delta^2 + (\tilde\beta_j^{(k-1)})^2} } \hat \beta_j\;, \quad j = 2,\dots,p \; 
\eeq
where we define $K \defeq   \tilde \lambda \s^2 /n$. Thus we have for each coefficient  a one-dimensional dynamic system independent of the other coordinates, which is easy to solve. Equation (\ref{Eq:ShrinkOrtho}) already indicates the shrinkage of the limit $\tilde\beta_j$ compared with the ML estimate $\hat \beta_j$. 
The stationary points of the sequence $\tilde\beta_j^{(k)}$ can be found by solving the equation
\beq \label{Eq:ShrinkOrtho2}
\tilde\beta_j \left(1 +  \frac{K}{\delta^2 + \tilde\beta_j^2}\right)  =    \hat \beta_j  \; .
\eeq

\begin{figure}[tb!]
\begin{center}
\ \\[-2.5cm]
\includegraphics[scale=0.4]{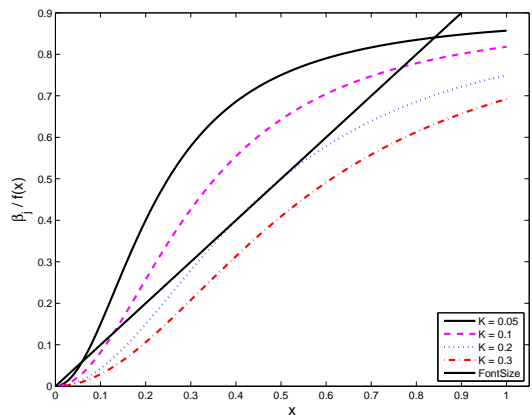}
\end{center}
\ \\[-2.5cm]
\caption{Function which determined the dynamic system (\ref{DynSyst}) for $\delta = 10^{-5}$ and $\hat \beta_j = 0.9$.} \label{fig:Dynamic}
\end{figure}

For the sake of notational convenience let's write  $x_k = \tilde\beta_j^{(k)}$. We thus study the dynamic system 
\beq\label{DynSyst}
x_k =  \frac{\hat \beta_j}{f(x_{k-1})}, \quad \mbox{with } \ f(x) = 1 + K(\delta^2 + x^2)^{-1} \; ,
\eeq
which is illustrated in Figure \ref{fig:Dynamic}. As long as $K > 8 \delta^2$ (which is essentially always the case) the function $xf(x)$ has two positive local extrema. The dynamic system (\ref{DynSyst}) has only one stationary point $x_I$ when the function value of its positive local minimum $x_*$ is larger than $\hat \beta_j$, that is
\beq \label{Cond:OneFP}
x_* f(x_*) > \hat \beta_j, \mbox{ with } x_*^2 = \frac K2 - \delta^2 + \frac 12 \sqrt{(K - 2\delta^2)^2 - 4\delta^2} \; .
\eeq
If $\delta \ll K$ this roughly means that $\hat \beta_j < 2 \sqrt K$. In that case it is easy to see that the only stationary point $x_I$ is attractive, and one has $x_k \rightarrow x_I \approx 0$ (see Figure~\ref{fig:Dynamic} for $K = 0.3$).

The other common situation occurs when the inequality  in (\ref{Cond:OneFP}) changes, that is when essentially $\hat \beta_j > 2 \sqrt K$. Then it holds that (\ref{Eq:ShrinkOrtho2}) has three solutions $x_I < x_{II} < x_{III}$.  Standard arguments from the theory of dynamical systems  show that $x_I$ and $x_{III}$ are attractive, that is for $x_1 <  x_{II}$ one has $x_k \rightarrow x_I$, otherwise if $x_1 > x_{II}$ then $x_k \rightarrow x_{III}$ (see Figure~\ref{fig:Dynamic} for $K = 0.1$ and $K = 0.05$). Note that $x_I, x_{II}$ and $x_{III}$ are the roots of a polynomial of third degree for which explicit formulas are available.

In the exceptional case where there are only two stationary points the dynamic is such that for $x_1 <  x_{II}$ one has again $x_k \rightarrow x_I$, but for 
$x_1 >  x_{II}$ one has  $x_k \rightarrow x_{II}$. Thus $x_{II}$ is a saddle point (see Figure~\ref{fig:Dynamic} for $K = 0.2$). 

Convergence of $x_k \rightarrow x_I$  can be interpreted  as $\tilde \beta_j = 0$, although 
$
0 < x_I \approx  \delta^2 \hat \beta_j / K
$.
However, numerically this is small enough to be indistinguishable from zero as long as $\delta$ is sufficiently small. Thus from a model selection perspective convergence towards $x_I$ indicates that a coefficient has been excluded, whereas the limit $x_{III}$ corresponds to regressors which have been included in the model. 
Furthermore equation (\ref{Eq:ShrinkOrtho2}) shows the amount of shrinkage that a regression coefficient suffers from AR. The larger  $x_{III}$ and the smaller $K$, the less shrinkage.

Two conditions have to be fulfilled that a regressor is selected by AR. Firstly the dynamical system of the component must have three fixed points, which corresponds to the condition that $K <  \hat \beta_j^2/4$. Secondly it is then necessary that $x_1 > x_{II}$.  
Remember that AR computes in its first step $\tilde{\boldsymbol\beta}^{(1)}$ by standard ridge regression, and therefore $
x_1 =  \hat \beta_j/(1 + K)$.
On the other hand a very good approximation of $x_{II}$ can be obtained by letting $\delta = 0$ in (\ref{DynSyst}) and then solving the corresponding stationary equation, which results in 
$$
x_{II} \approx \hat \beta_j / 2 - \sqrt{\hat \beta_j^2 / 4 - K} \;.
$$
As long as $K < 1$ it then always holds that $x_1 > x_{II}$, and it follows that the dynamic of AR under orthogonality is completely determined by the number of fixed points for each regressor. To summarize, under orthogonality AR becomes  a thresholding procedure where a regressor is selected in case of
\beq \label{Eq:Thresh2}
\hat\beta_j^2 > 8 \lambda \s^2 / n \;.
\eeq
%
%
Comparing conditions (\ref{Eq:Thresh1}) and (\ref{Eq:Thresh2}) then yields

{\prop  \label{Prop1}
Under orthogonality performing AR with $\tilde \lambda$ corresponds to minimizing (\ref{Eq:PenLogLik})
with $\lambda = 4 \tilde \lambda$.
}
\ \\[3mm]
{\bf Remark:} \emph{The result  holds under the condition that $K<1$, or equivalently that $\tilde \lambda < n / \s^2$.  In practice this seems to be no huge restriction. To give some examples, the penalties of AIC, BIC and mBIC are $\lambda = 2, \lambda = \log n$ and $\lambda = \log(n p^2 / 4)$, respectively. As long as $y$ is reasonably scaled the condition  $K<1$ will always apply.}

\subsection{Non-orthogonal case} \label{SubSec:General}

For the non-orthogonal case a full analysis of the dynamical system (\ref{Eq:ItProc3}) becomes way more complicated, because it cannot be reduced any longer to independent analysis for the individual coefficients. Instead of attempting to obtain analytic results we will focus here on illustrating the most important features of AR by presenting results from simulations. Before that we only want to mention that as a simple consequence of (\ref{Eq:ItProc3})  it always holds that
\beq
\left\lVert\tilde{\boldsymbol\beta}^{(k)}\right\lVert \le \left\lVert\left(\boldsymbol X^T \boldsymbol X\right)^{-1} \boldsymbol X^T \boldsymbol y\right\lVert,
\eeq
and thus the sequence of $\tilde { \boldsymbol \beta}^{(k)}$ remains bounded. However, it turns out that the mapping underlying the dynamic system $\tilde {\boldsymbol \beta}^{(k)}$ is usually not a contraction, and therefore theoretical convergence results are rather hard to obtain. In fact changing the initial value of the weights $\boldsymbol w^{(0)}$ can have some effect on the limit of $\tilde {\boldsymbol \beta}^{(k)}$, though usually the obtained solutions are not too different from each other.

\begin{figure}[tbh!]
\begin{minipage}{4cm} Initial value $\boldsymbol w^{(0)} = 1$  \\[-3cm]  \centerline{\includegraphics[scale=0.35]{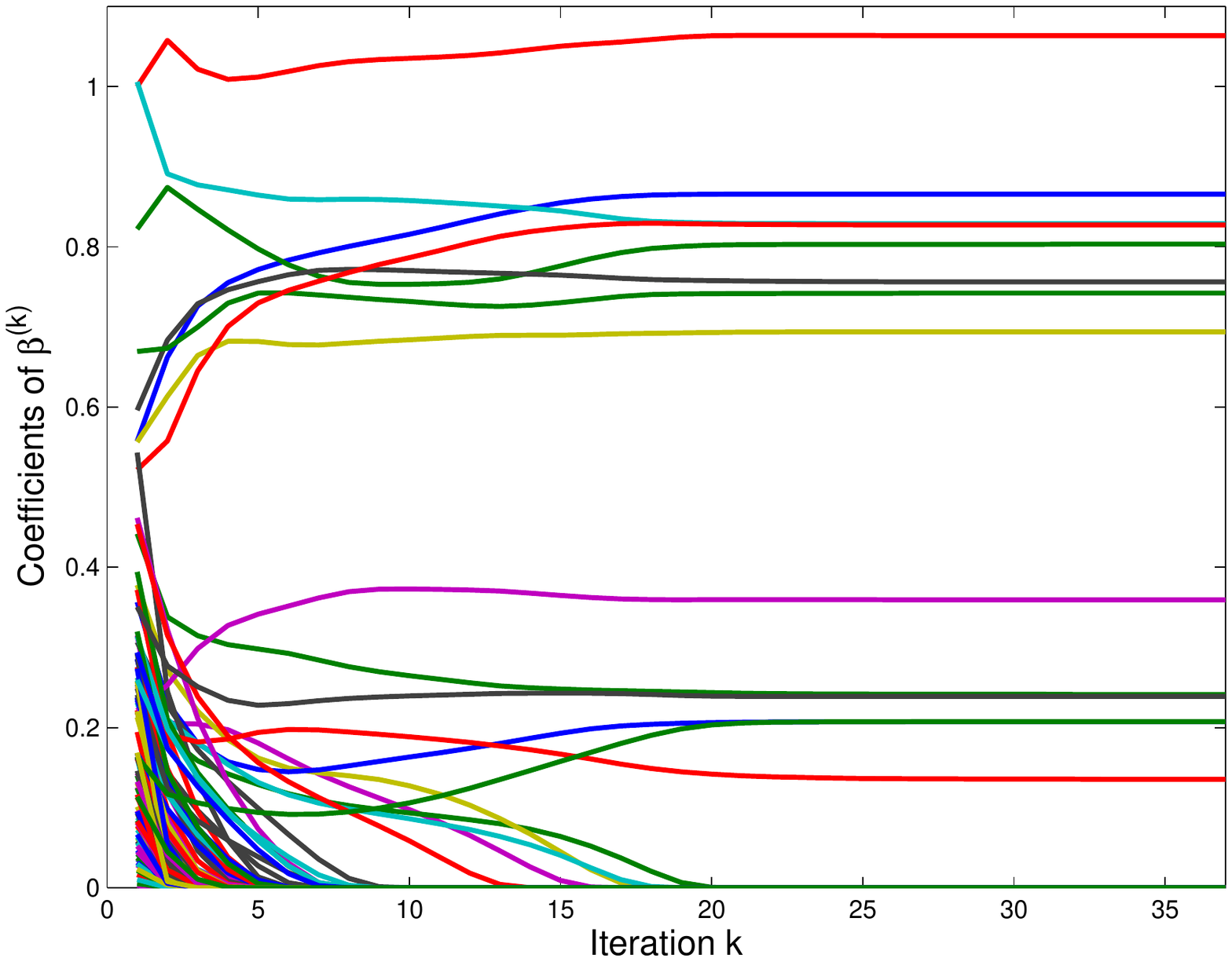}}
\end{minipage}
\hspace{2cm}
\begin{minipage}{4cm} Disturbed initial value  \\[-3cm] \centerline{\includegraphics[scale=0.35]{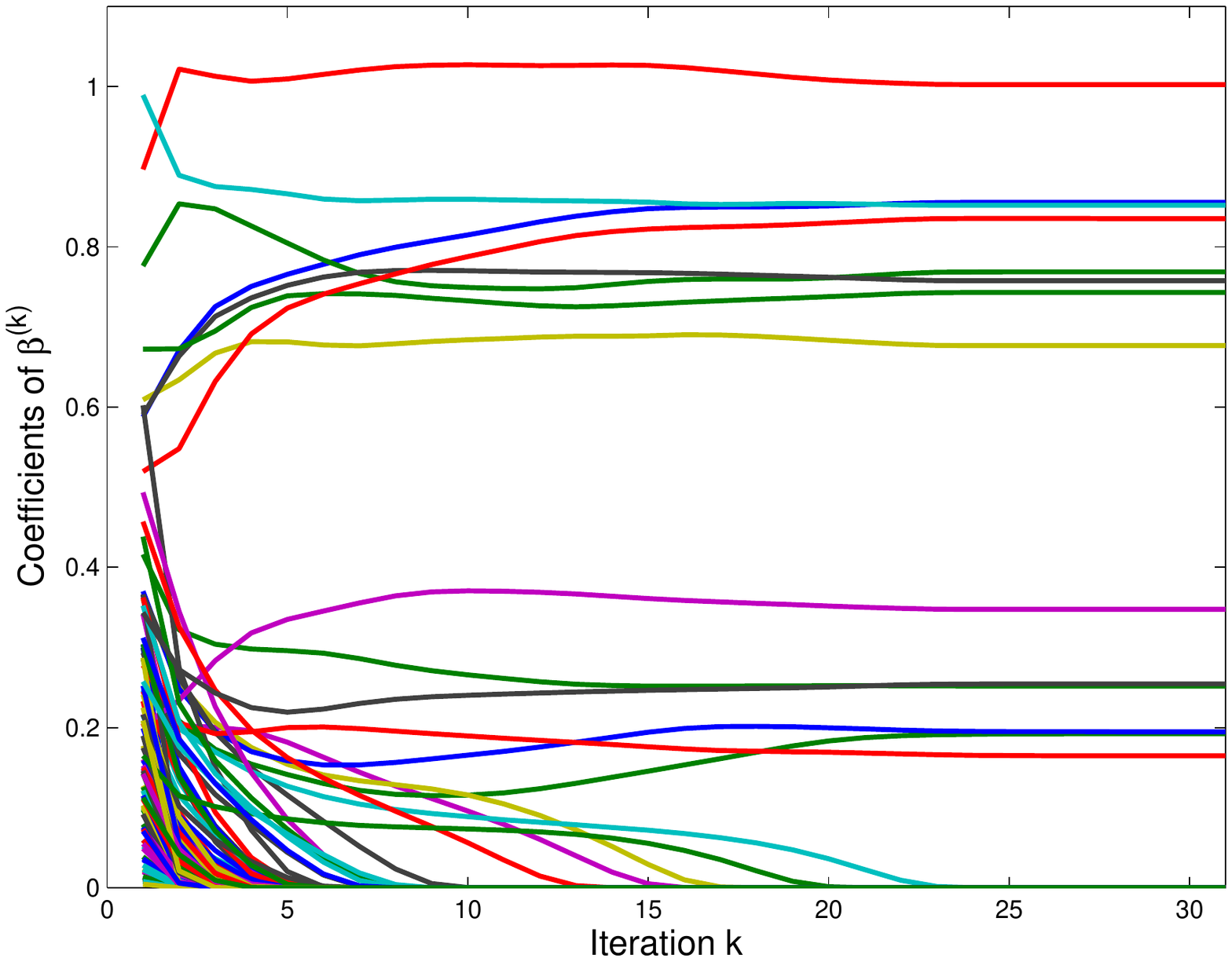}}
\end{minipage}
\\[-2.5cm]
\caption{Convergence of procedure for one simulated instance where the standard initial value is compared with a different choice of $\boldsymbol w^{(0)}$.}
\label{fig:Convergence}
\end{figure}

Figure~\ref{fig:Convergence} provides a typical example that illustrates the behavior of AR for the general linear case. We simulated one instance according to (\ref{Eq:LinModel}) with $p = n = 100$, where the correct model had $k^* = 24$ regressors. The first plot uses our standard initial value $w_j^{(0)} = 1$ for all components, whereas in the second plot the components of the initial value are randomly chosen between $1/2$ and $3/2$. The models resulting from the two starting points differ only in one regressor, where a true positive detected by the second model is substituted in the first model by a false positive. Otherwise both models contain the same non-zero coefficients, for which estimates can also slightly differ. For this instance trying further random initial values of  $w_j^{(0)} \sim U(0.5, 1.5)$ provided a third limiting model which added one false positive to the second model. Interestingly the model obtained with the second starting point which was doing  best in terms of misclassification had the largest BIC criterion (141.03), while the other two models had almost identical BIC criterion (140.06 and  140.07). In general our experience with simulations shows that although the limit of the AR procedure depends on the starting point, the different solutions obtained will have very similar values of the selection criterion that one  attempts to approximate. In fact the instability of solutions does not come as a surprise bearing in mind that variable selection based on information criteria is well known to suffer from instabilities with respect to small changes within the data \cite{Chen2014}.  

Note that any component of the sequence $\tilde {\boldsymbol \beta}^{(k)}$ which once has approached zero also remains there. 
This can be easily understood because for small $\tilde {\boldsymbol \beta}_j^{(k)}$ the corresponding weight $w_j$ becomes very large, and the matrix $\boldsymbol X^T \boldsymbol X + \tilde \lambda \s^2 \diag(\boldsymbol w)$  becomes essentially orthogonal with respect to the $j$-th component.
This mechanism of the procedure can be used to force some $\beta_j$ to stay in the model regardless of the penalty, simply by setting its corresponding initial weight $w_j$ to $0$. This might  be useful in practice if one would not like to perform model selection on a certain subset of regressors.
 The majority of coordinates converging to zero does so within less than 10 iterations, but there are some exceptions for which convergence to zero takes substantially longer. The instability of the AR model as a function of the initial values appears to depend mainly on the behavior within the first few iterations, where for the majority of coefficients it becomes clear whether they are selected or not.

\subsubsection{Correlated regressors}

Next we look at two very simple scenarios, where we study more systematically the behavior of AR  when regressors are correlated. To this end we consider correlation structures from compound symmetry and auto regressive models.
In both scenarios the  parameter $\rho$ varies between 0 and 0.8, where $\rho$ specifies pairwise correlation between neighboring covariates  for auto regression (Scenario 2), and pairwise correlation between all regressors for compound symmetry (Scenario 1). We consider with  $p = 15$ a relatively small number of regressors. This allows for a systematic examination of the performance of AR compared with all subset selection, which is for $p = 15$ still conveniently possible. For each scenario we simulate 500 traits for $n = 50$ individuals based on linear models with 5 regressors having nonzero coefficients. The effects are all chosen to be $\beta_j =  0.5$, which equals half of the predefined standard deviation $\sigma = 1$. Regressors entering the model were chosen to be  $j \in \{1,\dots,5\}$ for Scenario 1, and $j \in \{2,5,8,11,14\}$ for the second scenario. Selection based on BIC is compared with AR using parameter $\lambda = \log(n) / 4$, that is we use the  relationship $\lambda = 4 \tilde \lambda$  as suggested by Proposition \ref{Prop1}.

\begin{figure}[tbh!]
\begin{minipage}{4cm} Compound Symmetry  \\[-3cm]  \centerline{\includegraphics[scale=0.35]{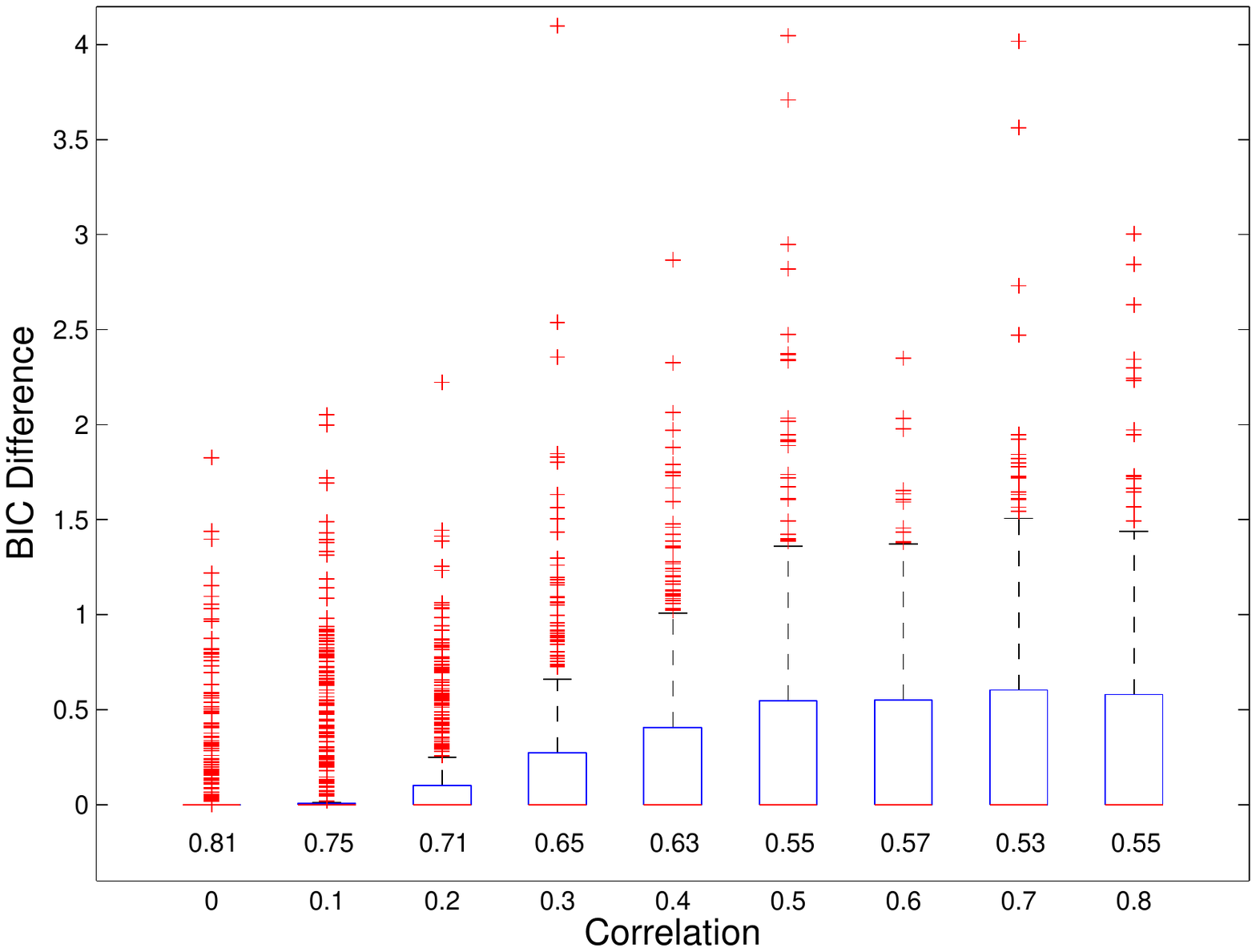}}
\end{minipage}
\hspace{2cm}
\begin{minipage}{4cm} Auto regression  \\[-3cm] \centerline{\includegraphics[scale=0.35]{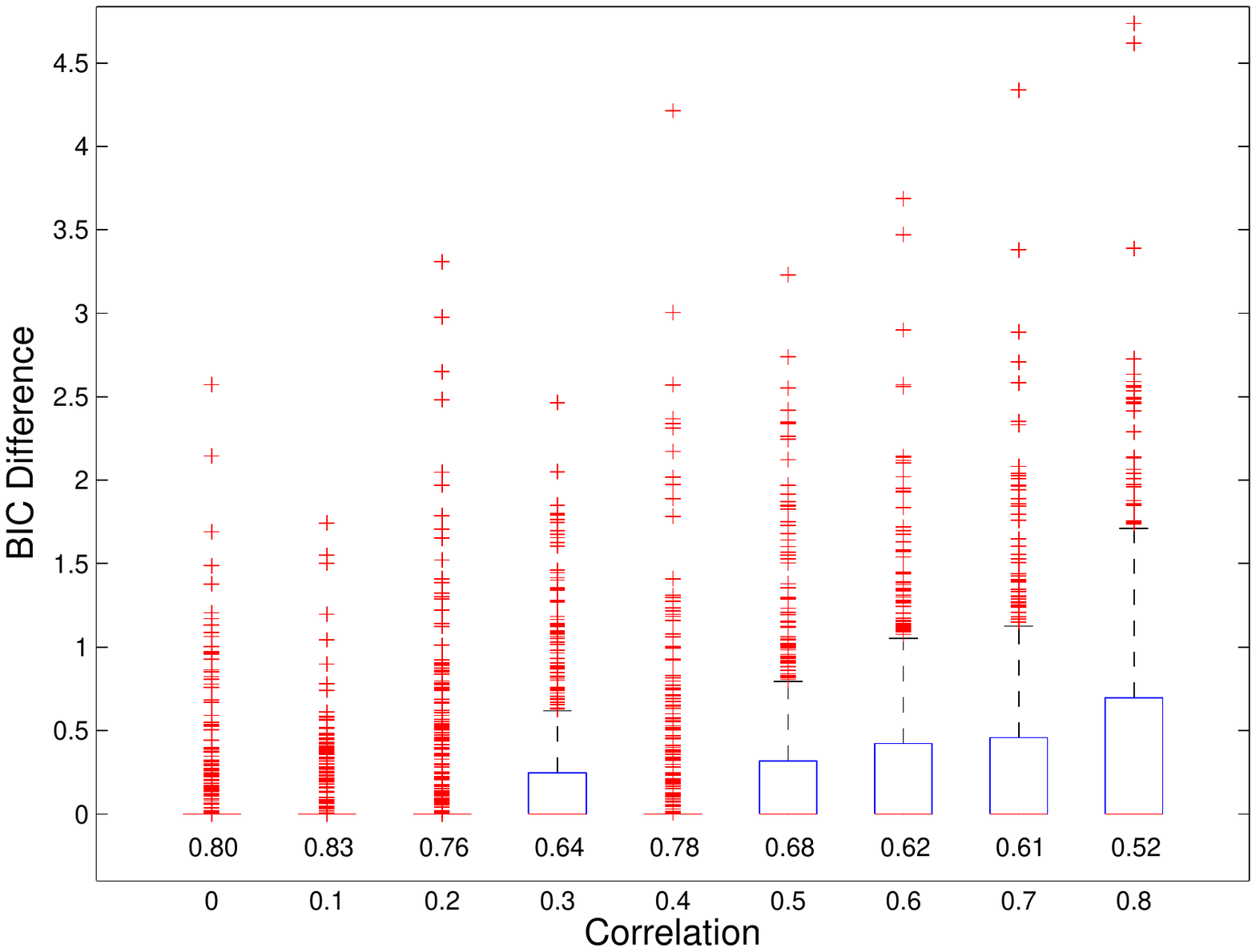}}
\end{minipage}
\label{fig:AllSubset_crit}
\\[-2.5cm]
\caption{Difference of BIC between  model obtained with AR and best model. The numbers below the boxplots give the relative frequency of simulation runs in which AR gave the optimal solution.}
\end{figure}

Figure \ref{fig:AllSubset_crit} illustrates to which extent AR yields the optimal model according to BIC. For small correlations AR gives in the majority of cases the same model as all subset selection, which starts to change only  for $\rho \geq 0.3$. In Scenario 2 AR yields more often the optimal model than in Scenario 1, when comparing results at the same level of pairwise correlation. Clearly a compound symmetry model provides in general more correlation between regressors than an autoregressive model, and we might conclude that AR differs increasingly from all subset selection the farther away one gets from orthogonality. 

\begin{table}[b!]
\caption{Comparison of the performance of all subset selection (BIC) with AR in terms of 
power, number of false positives (FP), false discovery rate (FDR) and number of misclassifications (Mis). For Scenario 1 the correlation (Corr) refers to pairwise correlation between all regressors, for Scenario 2 only for neighboring regressors. }  \label{Tab:procedures}
\begin{center}
\begin{tabular}{c|ll|ll|ll|ll}
 & \multicolumn{2}{c}{Power} & \multicolumn{2}{c}{FP} & \multicolumn{2}{c}{FDR} & \multicolumn{2}{c}{Mis} \\
Corr&BIC&AR&BIC&AR&BIC&AR&BIC&AR\\
\hline
\multicolumn{9}{l}{\bf Scenario 1:} \\
\hline
0.0 & 0.85 &0.83 &0.54 &0.52 &0.11 &0.10 &1.30 &1.39\\
0.1 & 0.81 &0.84 &0.61 &0.62 &0.11 &0.11 &1.54 &1.42\\
0.2 & 0.82 &0.86 &0.56 &0.51 &0.10 &0.09 &1.47 &1.23\\
0.3 & 0.79 &0.83 &0.65 &0.66 &0.13 &0.12 &1.71 &1.50\\
0.4 & 0.75 &0.80 &0.61 &0.62 &0.12 &0.12 &1.86 &1.60\\
0.5 & 0.68 &0.74 &0.79 &0.73 &0.17 &0.15 &2.38 &2.02\\
0.6 & 0.66 &0.72 &0.61 &0.65 &0.14 &0.14 &2.30 &2.04\\
0.7 & 0.56 &0.62 &0.71 &0.76 &0.19 &0.18 &2.90 &2.68\\
0.8 & 0.49 &0.54 &0.84 &0.87 &0.25 &0.24 &3.41 &3.19\\
\hline
\multicolumn{9}{l}{\bf Scenario 2:} \\
\hline
0.0 & 0.85 &0.83 &0.54 &0.47 &0.10 &0.10 &1.29 &1.35\\
0.1 & 0.86 &0.88 &0.55 &0.53 &0.10 &0.09 &1.24 &1.11\\
0.2 & 0.81 &0.80 &0.62 &0.54 &0.13 &0.11 &1.58 &1.52\\
0.3 & 0.67 &0.62 &0.70 &0.65 &0.19 &0.18 &2.37 &2.56\\
0.4 & 0.88 &0.88 &0.74 &0.72 &0.13 &0.13 &1.36 &1.32\\
0.5 & 0.81 &0.83 &0.84 &0.79 &0.17 &0.15 &1.81 &1.62\\
0.6 & 0.80 &0.80 &0.91 &0.93 &0.18 &0.18 &1.93 &1.94\\
0.7 & 0.72 &0.75 &1.00 &0.89 &0.21 &0.18 &2.41 &2.13\\
0.8 & 0.61 &0.65 &1.30 &1.24 &0.30 &0.28 &3.27 &3.00\\
\end{tabular}
\end{center}
\end{table}

  Interestingly from a statistical point of view AR seems to perform almost better than all subset selection based on BIC. For the majority of cases AR has less misclassifications than all subset selection (see Table \ref{Tab:procedures}). Specifically for Scenario 1 AR tends to have larger power to detect the correct regressors, while controlling the Type I error at a similar rate like BIC. On the other hand in Scenario 2 AR tends to give less Type I errors, while having similar power to BIC. In summary one might conclude that for $p < n$ (at least in these two scenarios) the choice of $\lambda = 4 \tilde \lambda$  from Proposition \ref{Prop1} worked quite well even in the non-orthogonal case.

\subsubsection{High-dimensional setting}

A large number of recent statistical applications are confronted with the challenging task of model selection when $p > n$. Here we perform simulations  under the assumption that regressors are independent normally distributed variables. Sample size was fixed with $n=100$, while for the growing number of potential regressors we considered  $p \in\{ 100, 250, 500, 1000, 2500, 5000, 10000 \}$. For each setting 1000 models of size $k^* = 24$ were simulated from (\ref{Eq:LinModel}), with normally distributed random effect sizes $\beta_j \sim N(0, 0.5), j \in \{1,\dots,24\},$ and again an error standard deviation of $\sigma = 1$. Keeping $k^*$ fixed gives with growing $p$  an increasingly sparse situation. Hence the model selection criterion mBIC is more appropriate than BIC (see \cite{BZG08}), but here we are mainly interested in studying the properties of AR and will therefore show results for both criteria.

\begin{figure}[tbh!]
\begin{minipage}{5cm}  Power \\[6mm]  \centerline{\includegraphics[width = 1.7cm, bb = 200 200 350 550]{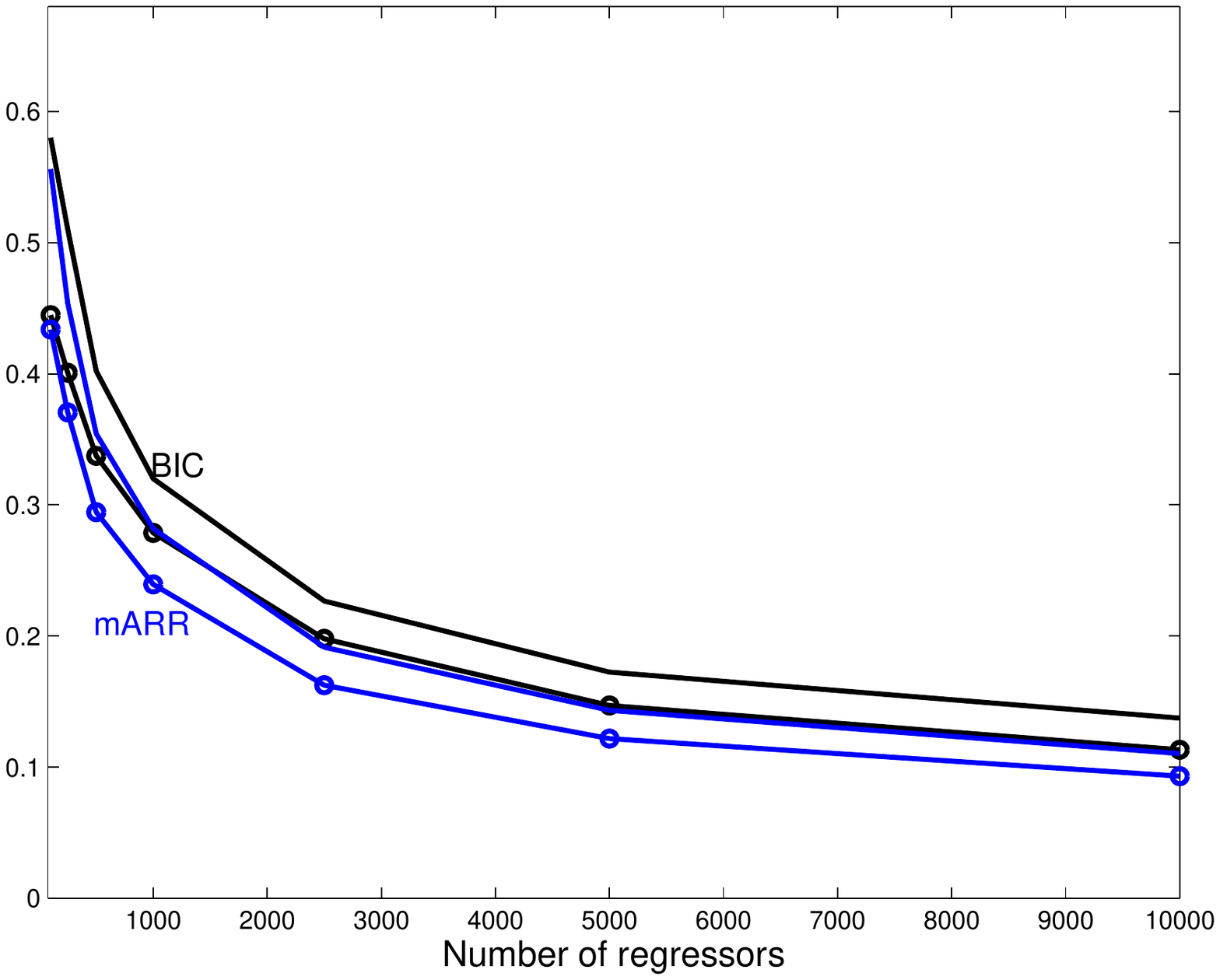}}
\end{minipage}
\hspace{1.4cm}
\begin{minipage}{5cm} Average number of FP \\[6mm]  \centerline{\includegraphics[width = 1.7cm, bb = 200 200 350 550]{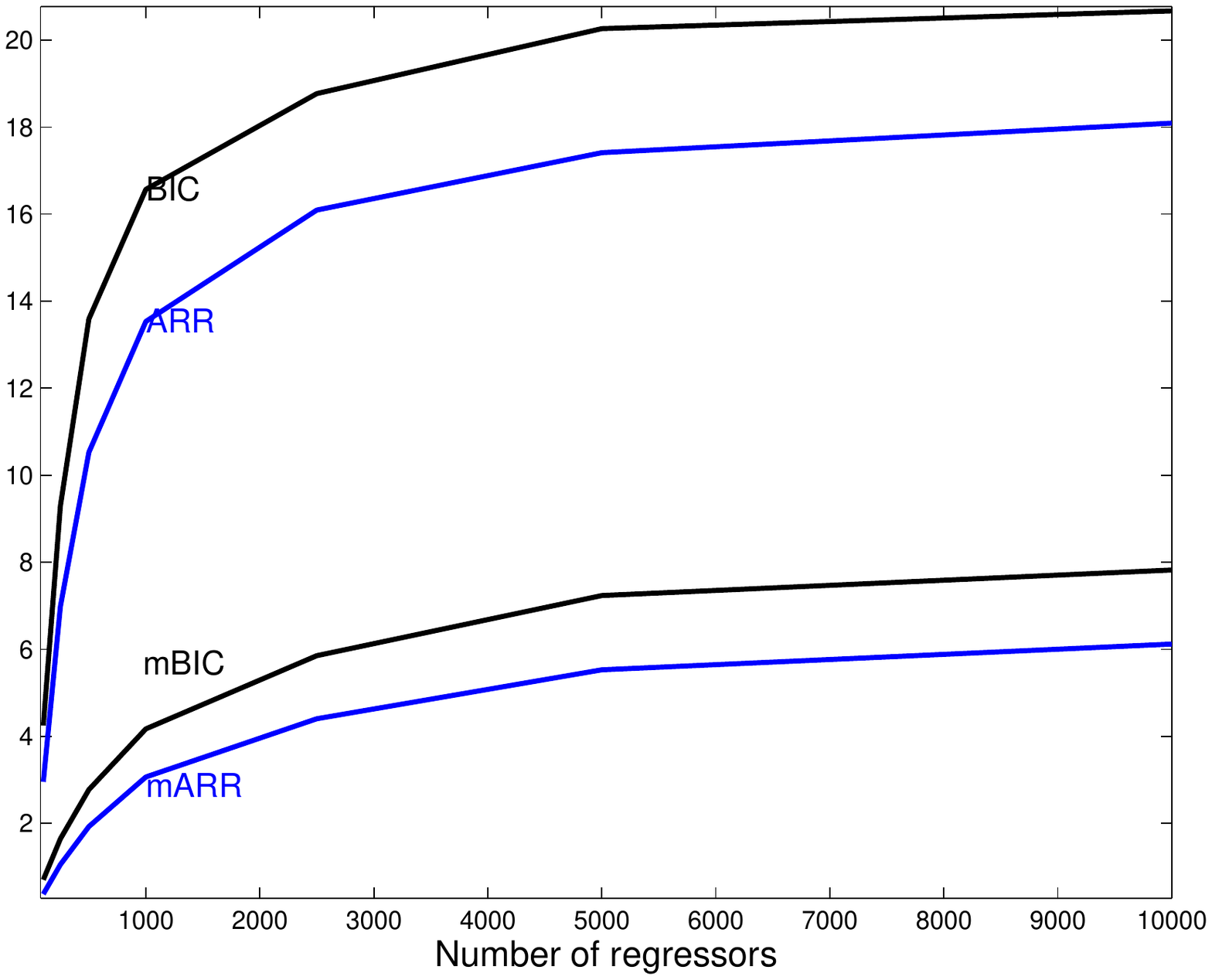}}
\end{minipage}

\ \\

\begin{minipage}{5cm}  Number of Misclassifications  \\[6mm]   \centerline{\includegraphics[width = 1.7cm, bb = 200 200 350 550]{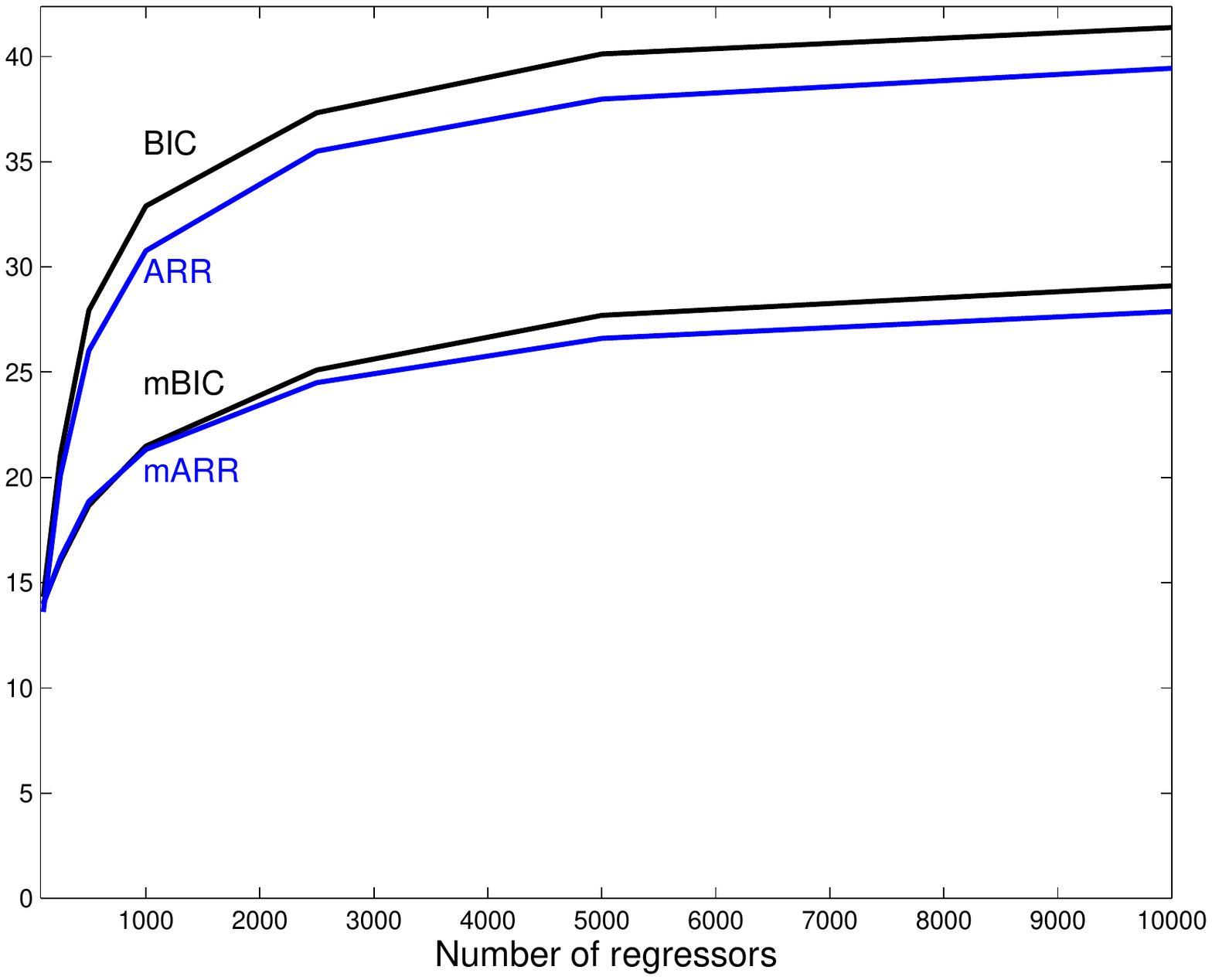}}
\end{minipage}
\hspace{1.4cm}
\begin{minipage}{5cm} FDR   \\[6mm]  \centerline{\includegraphics[width = 1.7cm, bb = 200 200 350 550]{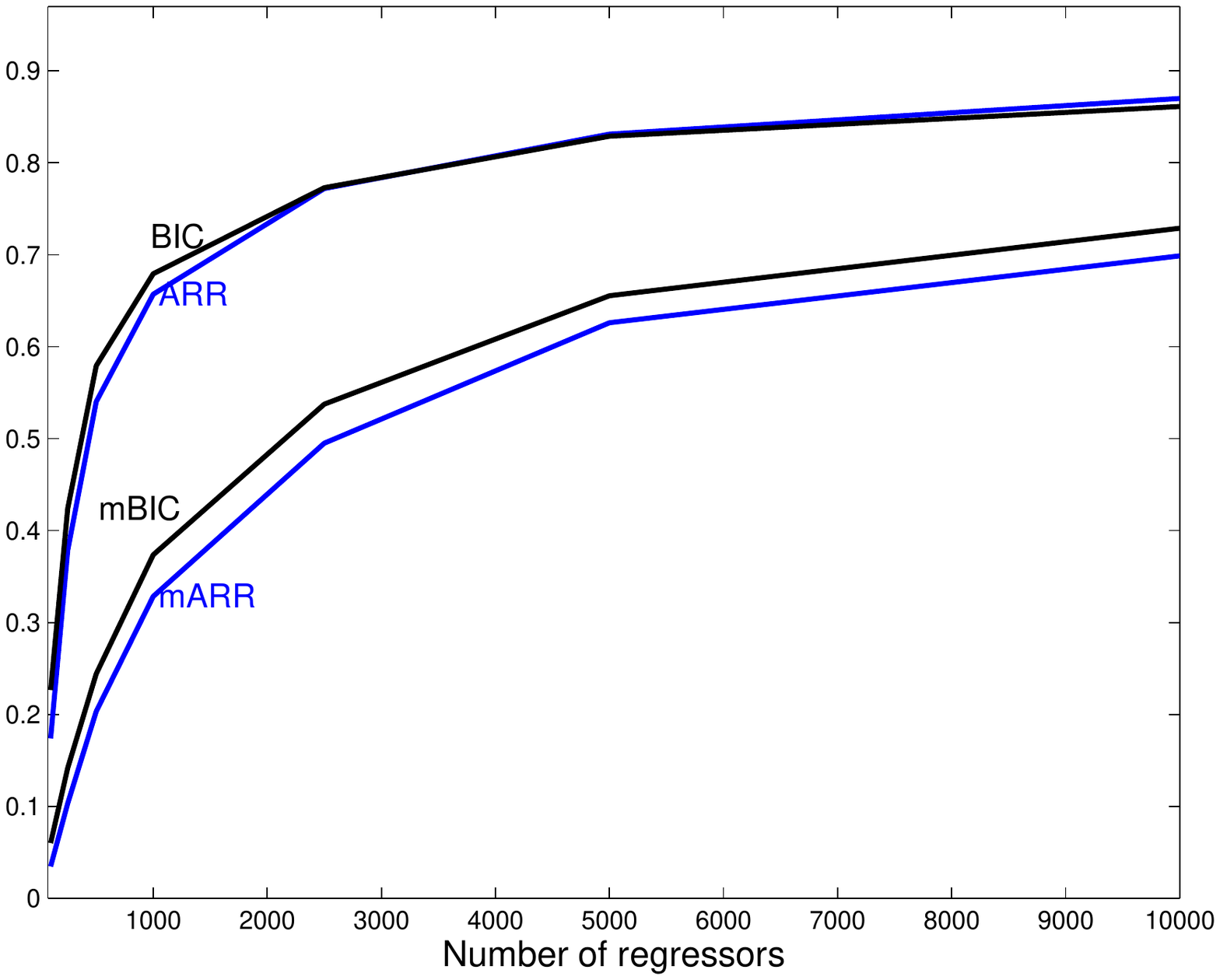}}
\end{minipage}
\caption{Comparison of model selection performance based on some stepwise selection procedure for BIC and mBIC with the corresponding AR procedures.  The four panels show the average over 1000 simulation runs of power, number of false positives, number of misclassifications and false discovery rate as a function of the total number of potential regressors $p$. Data were simulated under a model with $k = 24$ regressors.
} \label{Fig:p_larger_n}
 \end{figure}

We want to compare the performance of variable selection using simple stepwise search strategies for the two information criteria BIC and mBIC with their respective AR procedures. Our stepwise procedure is fairly simple. It starts with a model including the best 40 regressors according to marginal test statistics. Then greedy backward elimination is performed all the way down to a model of size one. That model along the way which minimizes the criterion in question is then considered as the starting point for some final greedy forward selection which is performed till no more improvement of the criterion is obtained. For the AR procedure we use again the relationship $\lambda = 4 \tilde \lambda$  from Proposition \ref{Prop1}.  Before applying AR the top 100 regressors were preselected based on marginal tests, which noticeably improved the performance of AR.

We start with discussing Figure \ref{Fig:p_larger_n}, which compares classification characteristics of the four procedures. Only for $n = p = 100$ BIC and mBIC are comparable in terms of misclassification.  With growing $p$  BIC produces exceedingly more false positives than mBIC, which cannot be compensated by the relatively mild gain in power. Both for BIC and mBIC the AR procedure is more conservative than the corresponding stepwise selection procedure, which means that it is less powerful, but produces also less false positives. Interestingly for both criteria AR produces less misclassifications than stepwise selection.

\begin{figure}[tbh!]
\begin{minipage}{5cm}  BIC \\[6mm]  \centerline{\includegraphics[width = 1.7cm, bb = 200 200 350 550]{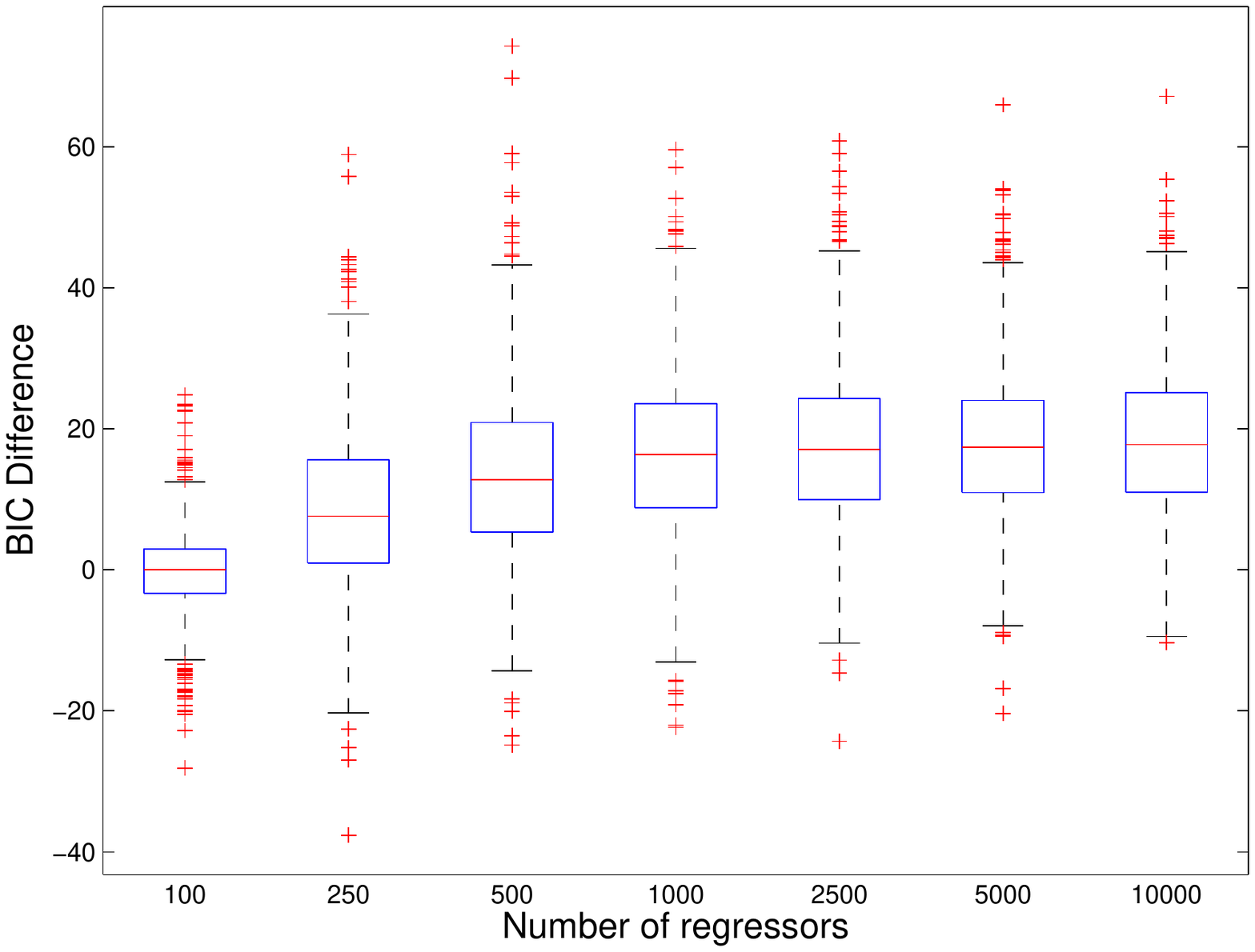}}
\end{minipage}
\hspace{1.4cm}
\begin{minipage}{5cm} mBIC \\[6mm]  \centerline{\includegraphics[width = 1.7cm, bb = 200 200 350 550]{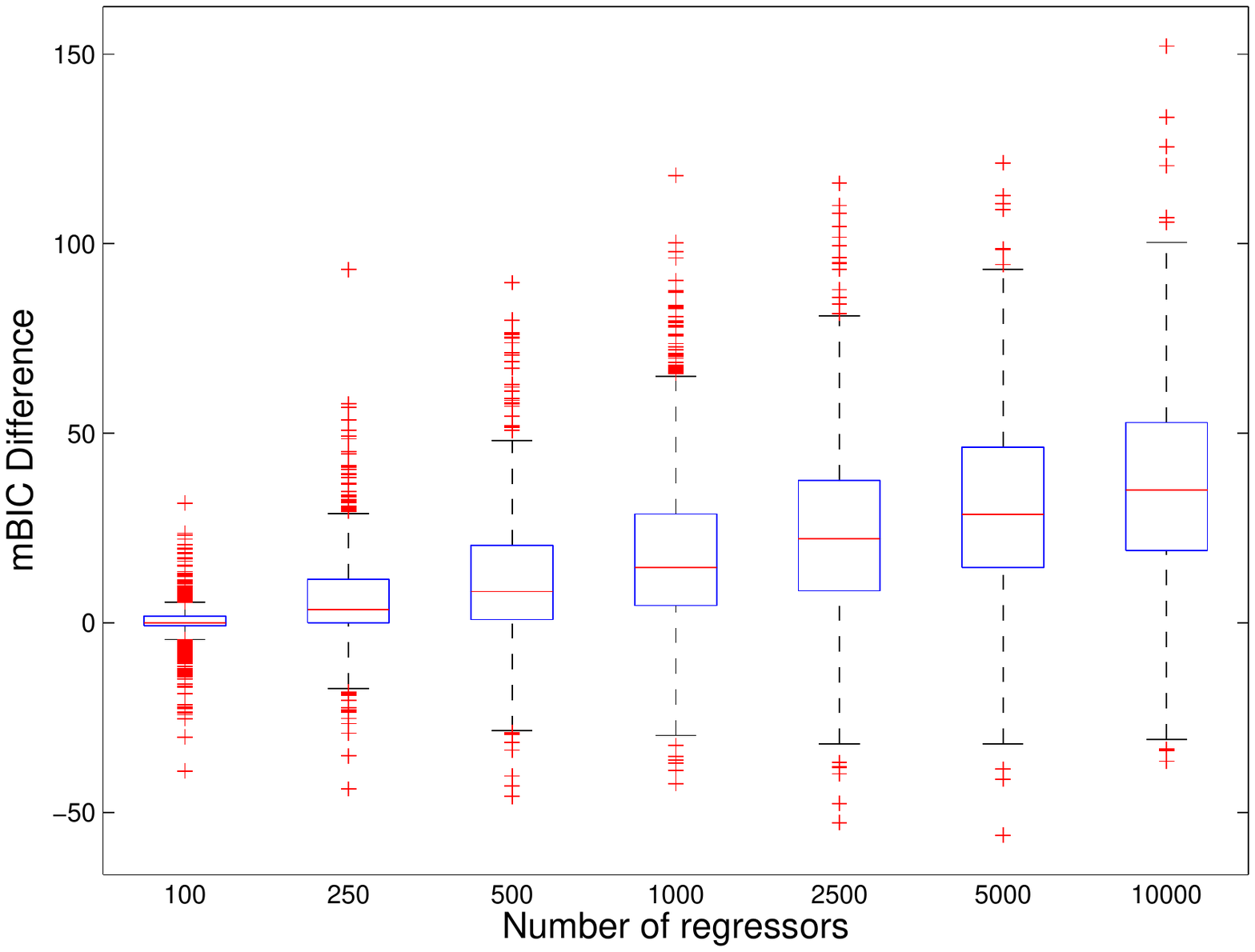}}
\end{minipage}
\caption{ Boxplots of differences between values of selection criteria for models obtained with stepwise search strategy and with AR. The first panel shows results for BIC, the second panel for mBIC. Results are based on the same data as Figure \ref{Fig:p_larger_n}.
} \label{Fig:DiffCrit}
 \end{figure}

Looking again at the differences of criteria for models obtained with stepwise selection and with AR, one can see that for $p$ getting larger AR tends to give models with  larger values of the criterion than stepwise selection. However, even for the largest $p$ there are at least some instances where AR gives better models according to each criterion than stepwise selection. For $p = n = 100$ AR and stepwise selection perform more or less identical, where the median of differences is almost exactly at 0. In case of BIC the median of differences increases with $p$ till $p=1000$ and then remains constant, whereas for mBIC the median of differences continues to grow also for larger values of $p$. It is interesting to observe that although for $p > n$ AR does usually not manage to find those models that minimize the information criterion, it outperforms the corresponding stepwise selection procedure with respect to misclassification. 

The fact that the AR procedure is for $p > n$ more conservative than stepwise selection gives rise to the question whether the relationship $\lambda = 4 \tilde \lambda$  from Proposition \ref{Prop1} is still correct, or whether one would rather have to use in that situation more relaxed penalties  to compensate for shrinkage. Our simulation results did not provide a definite answer to this question, but we will see in the next section on generalized linear models that in principal it is easy to obtain solutions of AR for a whole range of $\tilde \lambda$ values, among which one can then choose the model which minimizes the original $L_0$ penalty with parameter $\lambda$.

\section{Further Applications} \label{Sec:Ex}

In this section, we consider two more applications of the adaptive ridge approach in order to illustrate its usefulness beyond linear regression. We first discuss in Section~\ref{sec:GenLin} two particular cases of generalized linear models,  Poisson regression and logistic regression. As the weighted ridge problem associated with these two models has no closed-form solution, we rely upon the Newton-Raphson adaptive ridge version (\ref{eq:NRAR}) of our algorithm to solve the corresponding optimization problems. Afterwards we reconsider in Section~\ref{Subsec:Seg} the least squares segmentation problem  for which AR was first introduced in \cite{rippe2012vizualization},  but we improve on the   original publication by deriving explicit recursive formulas for solving the weighted ridge problem rather than relying on (sparse) LU decompositions. As a result, our approach is much faster than the original one.

\subsection{Generalized linear model}\label{sec:GenLin}

\subsubsection{Poisson regression}\label{sec:poisson}

To illustrate how to apply AR in the context of generalized linear models we will discuss Poisson regression and logistic regression models. We start with the classical Poisson regression problem $y_i\sim \mathcal{P}(\mu_i(\beta))$ where  $\mu_i(\boldsymbol \beta)=\exp(\boldsymbol X_i\beta)$ with $\boldsymbol y,\mu \in \R^n$, $\boldsymbol X \in \R^{n \times p}$, and $\boldsymbol \beta \in \R^p$. In order to maximize the $L_0$ penalized log-likelihood of the problem, we introduce for any penalty $\lambda \geq 0 $ and weight vector $\boldsymbol w\in \R^p$ the following weighted ridge penalized log-likelihood:
\beq\label{eq:penlik_poisson}
\ell(\boldsymbol \beta; \lambda, \boldsymbol w)=\text{const.} + \boldsymbol  \beta^T \boldsymbol  X^T \boldsymbol  y -\boldsymbol  u^T \boldsymbol \mu(\boldsymbol \beta) -\frac{1}{2}\lambda \boldsymbol  \beta^T \diag(\boldsymbol w) \boldsymbol  \beta \; ,
\eeq
where $\mathbf{u} \in \R^n$ is an all-one column-vector. For given $\lambda\geq 0$ we want to maximize this quantity using the Newton-Raphson algorithm. Simple computations give the first two derivatives of $\ell(\boldsymbol \beta; \lambda, \boldsymbol w)$,
\beq
\nabla \ell(\boldsymbol \beta; \lambda, \boldsymbol w)=\boldsymbol X^T \boldsymbol y - \boldsymbol X^T \boldsymbol  \mu(\boldsymbol \beta) -  \lambda   \diag(\boldsymbol  w) \boldsymbol  \beta ;
\eeq
\beq
\text{Hess}\,\ell(\boldsymbol \beta; \lambda, \boldsymbol w)= - \boldsymbol X^T \diag(\boldsymbol \mu(\boldsymbol \beta)) \boldsymbol X - \lambda \diag(\boldsymbol w).
\eeq
Maximizing Eq.~(\ref{eq:penlik_poisson}) can therefore be done iteratively using the following update for $\boldsymbol \beta$:
\beq
\boldsymbol \beta \leftarrow \boldsymbol \beta-\text{Hess}\,\ell(\boldsymbol \beta; \lambda, \boldsymbol w)^{-1} \nabla \ell(\boldsymbol \beta; \lambda, \boldsymbol w).
\eeq
The AR procedure also requires to update the weights $\boldsymbol w$ according to (\ref{eq:weightsL0}).
Combining both updates leads to the computationally efficient procedure whose R-code reads as follows: \\[5mm]


\begin{center}
\begin{minipage}{0.5\textwidth}
\begin{verbatim}
w=rep(1.0,p); beta=rep(0,p);
for (iter in 1:itermax) {
  mu=exp(X%*%beta)[,1]
  A=t(X)%*%diag(mu)%*%X+lambda+diag(w)
  b=t(X)%*%(y-mu)-lambda*w*beta
  beta=beta+solve(A,b)[,1]
  w=1.0/(beta^2+delta^2)
}
\end{verbatim}
\end{minipage}
\end{center}
\ \\[5mm]
We want to point our that the resulting code is quite compact and extremely easy to understand and to implement, which is in stark contrast to the available LASSO implementation of the same problem \citep{friedman2010regularization} which uses a rather delicate coordinate descent algorithm.

\begin{figure}[tb!]
\begin{center}
\includegraphics[width=0.6\textwidth]{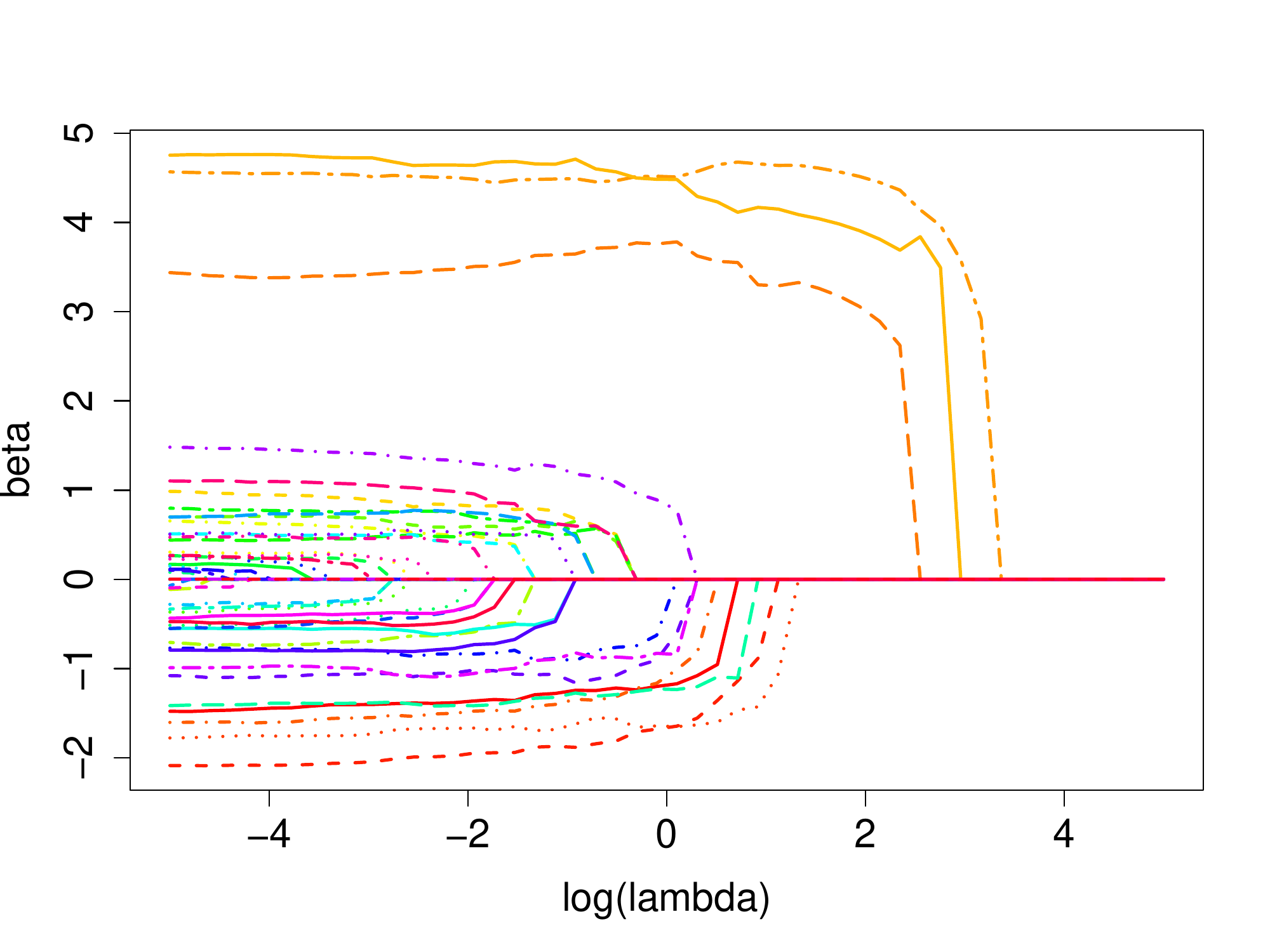}
\end{center}
\caption{Example of a full regularization path for $L_0$ adaptive Ridge Poisson regression with $n=300$, $p=50$, and $\boldsymbol \beta^*=0$ except for the first $k=10$ coordinates. The covariates $X_{i,j}$ and non-zero coefficients $\beta_j$ are independently drawn from normal random variables according to $X_{i,j} \sim \mathcal{N}(0,0.1^2)$ and  $\beta_j  \sim \mathcal{N}(0,1.5^2)$.}\label{fig:regpath_poisson}
\end{figure}

Like in case of LASSO it is also possible for AR to take advantage of a warm start of the algorithm to obtain the full regularization path of the problem (see Figure~\ref{fig:regpath_poisson}). For that purpose, we start with a near null penalty $\lambda$, and then increase the value of the penalty using for each new penalty the previously computed weight vector $w$ and parameter $\beta$ as starting points. Obtaining the full regularization path  is of particular importance in case of GLM because we do not have any theoretical results like Proposition \ref{Prop1} telling us which $\tilde \lambda$ of AR corresponds to the $\lambda$ of a given selection criterion. Interestingly, in case of Poisson regression and the BIC criterion it turns out that like in case of linear regression the factor 4 works really well.

\begin{figure}[tb!]
\begin{tabular}{cc}
\includegraphics[width=0.5\textwidth]{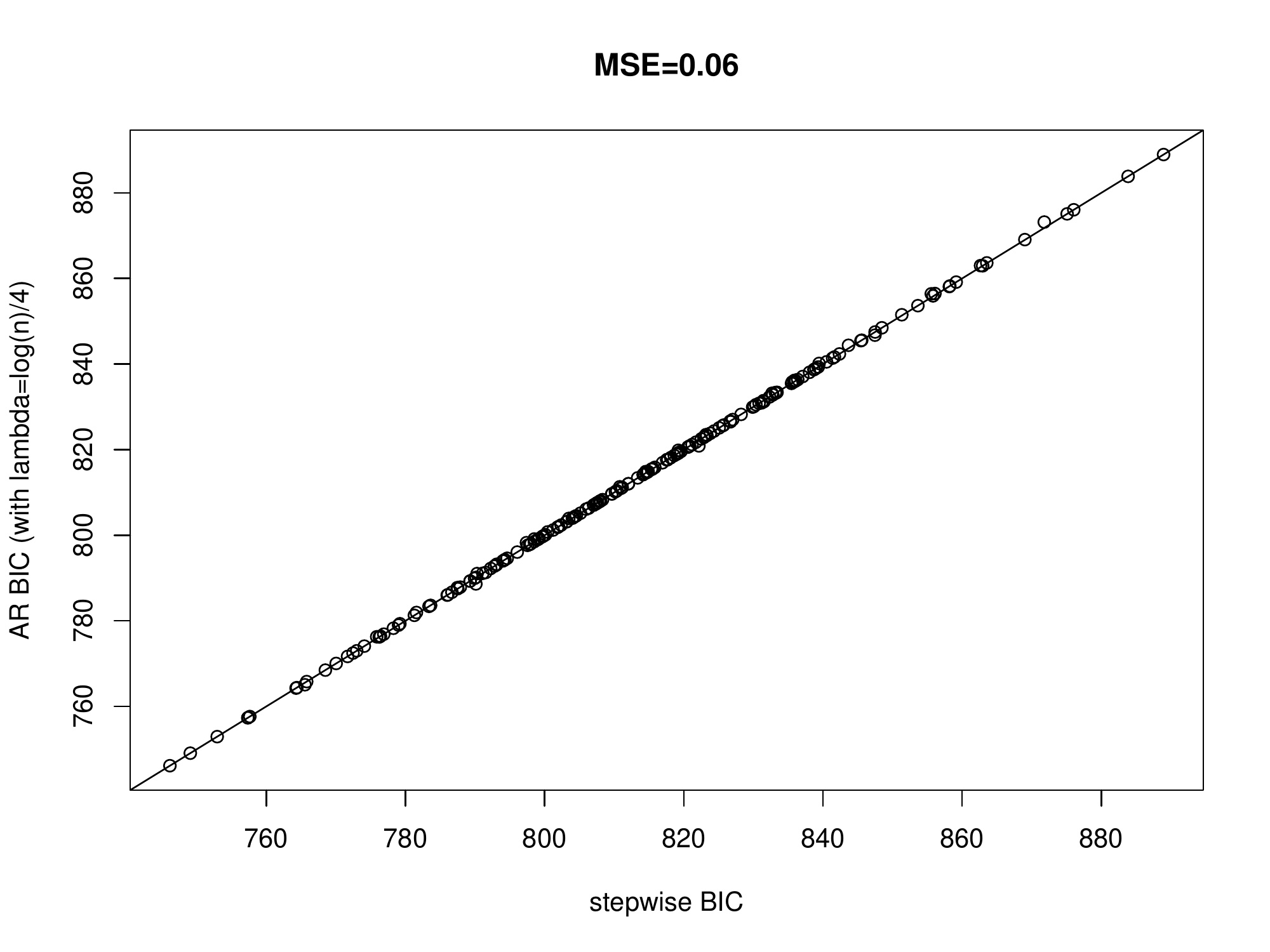}&
\includegraphics[width=0.5\textwidth]{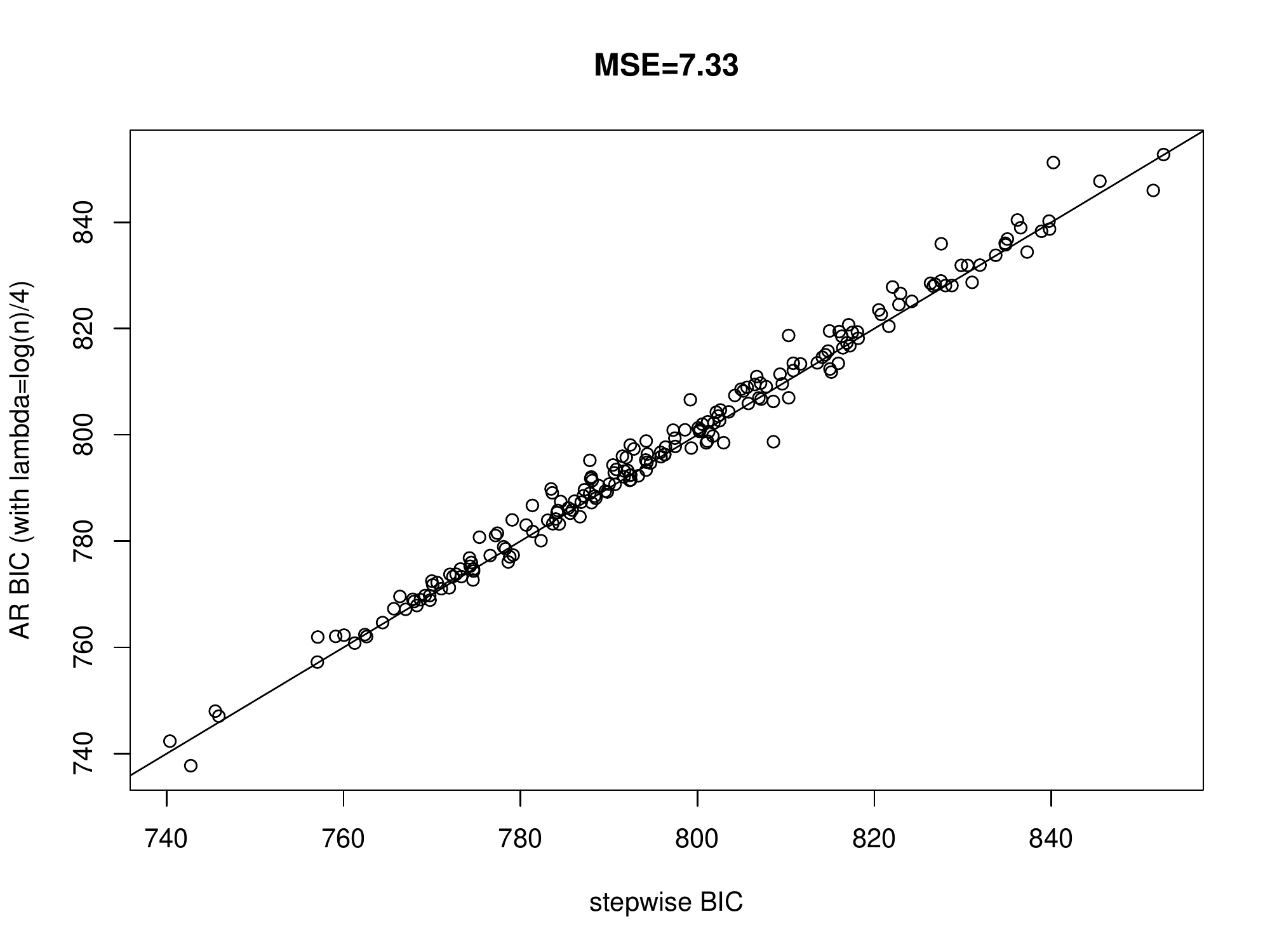}\\
(a)&(b)\\
\end{tabular}
\caption{Comparison of BIC criteria obtained with stepwise search and with AR for Poisson regression. Panel (a):  $n=300$, $p=50$, $k=10$;  Panel (b):  $n=300$, $p=500$, $k=25$,
 where $n$ is the sample size,  $p$ the total number of regressors and $k$ the size of the data generating model.
 } \label{fig:poisson}
\end{figure}

In Figure~\ref{fig:poisson} we compare AR Poisson regression with $\lambda=\log(n)/4$ to the standard stepwise selection procedure based on BIC. Two simulation scenarios are considered, the first one with $p=50$, the second one with $p = 500$, where both scenarios use a sample size of $n=300$. Count data were simulated from Poisson regression models of size $k = 10$ and  $k = 25$, respectively. 
 The covariates $X_{i,j}$ and the non-zero coefficients $\beta_j$ were independently drawn from normal random variables according to $X_{i,j} \sim \mathcal{N}(0,0.1^2)$ and  $\beta_j  \sim \mathcal{N}(0,1.5^2)$. 

For the first scenario with $p < n$ AR and stepwise selection give almost identical results, which is quantified by the extremely small mean squared error (MSE) of the difference between obtained BIC values. This illustrates on the one hand that the Newton-Raphson AR procedure (\ref{eq:NRAR}) works really well, and on the other hand that $\lambda=\log(n)/4$ is a perfect choice in this setting. In the high-dimensional setting with $p=500$ both methods still give very similar results, but with more distinct differences. Note however that differences go in both directions, and there is no clear trend observable that  stepwise procedures would give better results than AR.  

\subsubsection{Logistic regression}\label{sec:logit}

The classical binary logistic regression model is of the form $y_i\sim \mathcal{B}(\pi_i(\boldsymbol \beta))$ where  $\pi_i(\boldsymbol \beta)=1/(1+\exp(-\boldsymbol X_i\boldsymbol \beta))$
 with $\boldsymbol y\in\{0,1\}^n$, $\boldsymbol \pi \in [0,1]^n$, $\boldsymbol X \in \R^{n \times p}$, and $\boldsymbol \beta \in \R^p$. Just like in the case of Poisson regression we introduce for any penalty $\lambda \geq 0 $ and weight vector $w\in \R^p$ the weighted Ridge penalized log-likelihood:
\beq\label{eq:penlik_logistic}
\ell(\boldsymbol \beta; \lambda, \boldsymbol w)= \boldsymbol u^T \left\{(1-\boldsymbol y)\log(1-\boldsymbol \pi(\boldsymbol \beta)) + y \log(\boldsymbol \pi(\boldsymbol \beta))\right\} -\frac{1}{2}\lambda\boldsymbol  \beta^T \diag(\boldsymbol w) \boldsymbol \beta \; ,
\eeq
where $\mathbf{u} \in \R^n$ is an all-one column-vector. The Newton-Raphson AR (\ref{eq:NRAR}) needs again the two first derivatives of that penalized likelihood function, which are 
\beq
\nabla \ell(\boldsymbol \beta; \lambda, \boldsymbol w)=\boldsymbol X^T ((1-\boldsymbol \pi(\boldsymbol \beta))\boldsymbol y-(1-\boldsymbol y)\boldsymbol \pi(\boldsymbol \beta))-  \lambda   \diag(\boldsymbol w) \boldsymbol \beta;
\eeq
\beq
\text{Hess}\,\ell(\boldsymbol \beta; \lambda, \boldsymbol w)= -\boldsymbol X^T \diag(\boldsymbol \pi(\boldsymbol \beta)(1-\boldsymbol \pi(\boldsymbol \beta)) \boldsymbol X - \lambda \diag(\boldsymbol w).
\eeq

Like for Poisson regression we compared AR logistic regression to a standard stepwise selection procedure based on BIC. 
We present simulation results for a scenario which is similar to the first scenario for Poisson regression, with $p = 50$, $n = 300$ and $k = 10$. Again both the covariates $X_{i,j}$ and the non-zero coefficients $\beta_j$ were independently drawn from normal random variables according to $X_{i,j} \sim \mathcal{N}(0,0.1^2)$ and  $\beta_j  \sim \mathcal{N}(0,3.5^2)$. 
  
\begin{figure}[tb!]
\begin{tabular}{cc}
\includegraphics[width=0.5\textwidth]{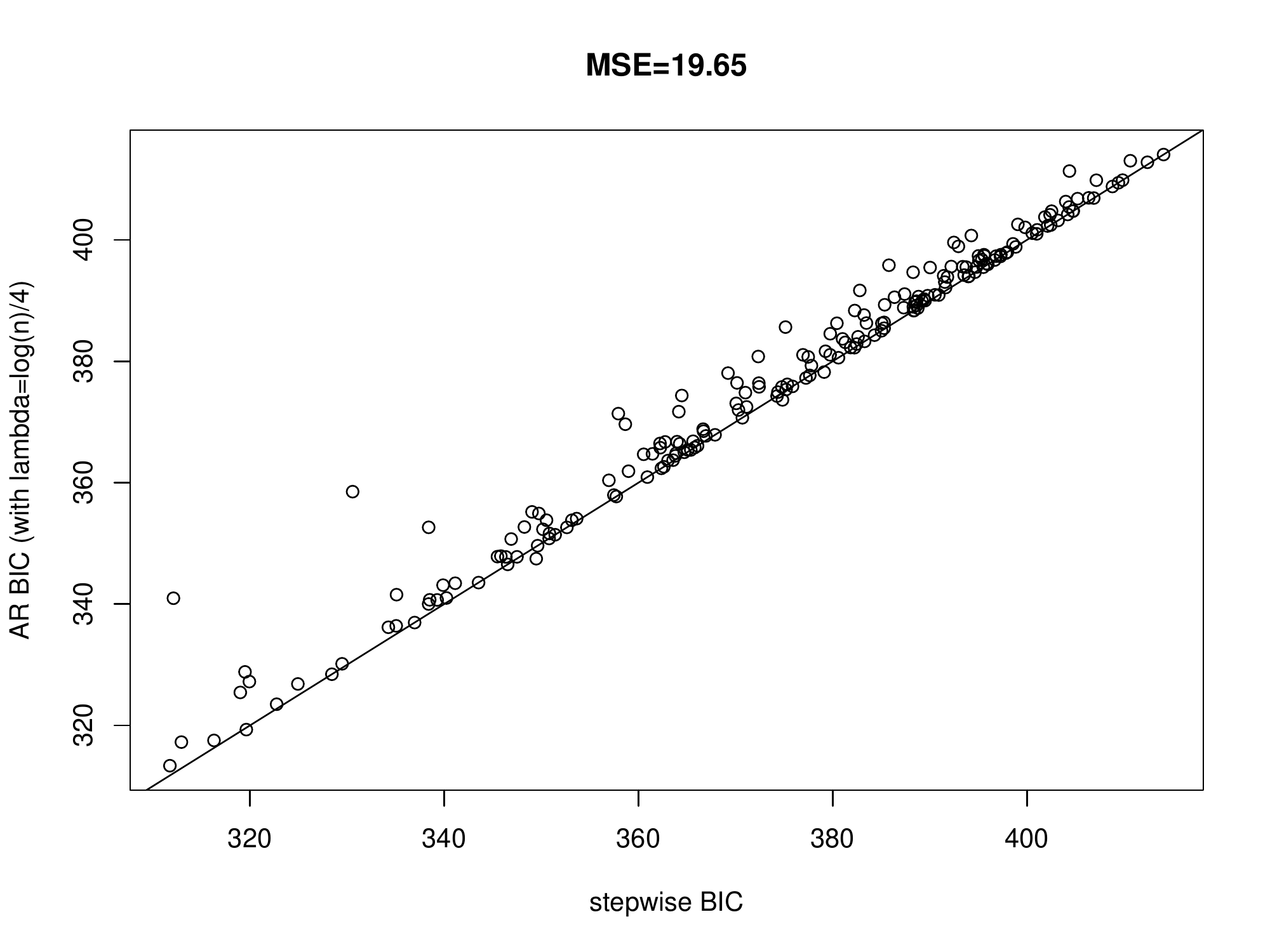}&
\includegraphics[width=0.5\textwidth]{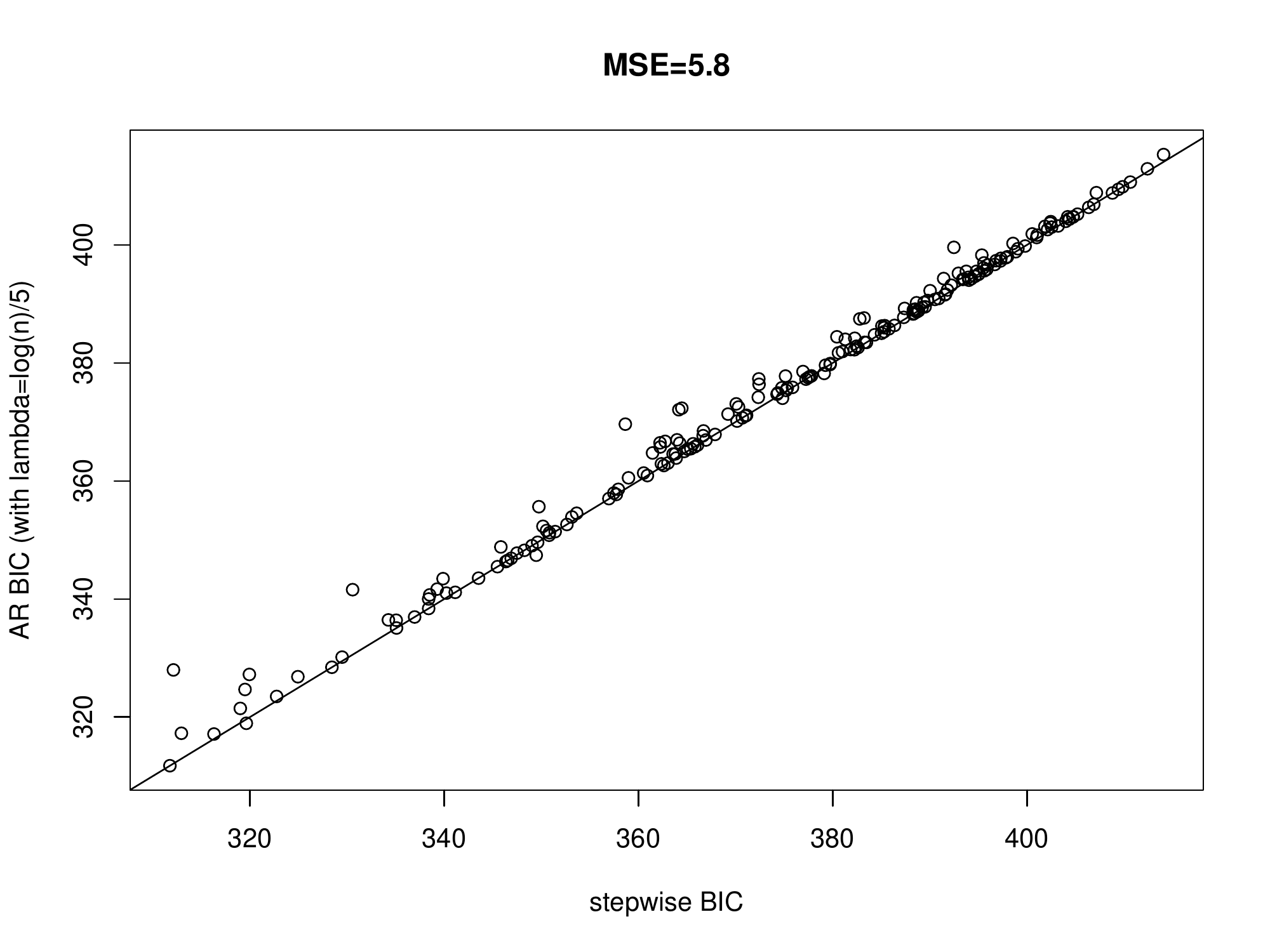}\\
(a)&(b)\\
\end{tabular}
\caption{{Comparison of BIC criteria obtained with stepwise search and with AR for logistic regression with $n=300$, $p=50$, and $k=10$.  In Panel (a) AR is performed with $\lambda=\log(n)/4$, in Panel (b) with $\lambda=\log(n)/5$.}
}\label{fig=logistic}
\end{figure}

Figure~\ref{fig=logistic} illustrates that for logistic regression the relationship $\lambda=\log(n)/4$ no longer gives the best results, but 
that the slightly smaller penalty of $\lambda=\log(n)/5$ performs better for this scenario.  No more improvement of MSE was observed by further decreasing the penalty $\lambda$ (data not shown).  For $\lambda=\log(n)/5$ AR gives quite similar results to stepwise selection, but the agreement is not as strong as in the corresponding scenario of Poisson regression, and if there are differences then in the majority of cases AR tends to give larger values of BIC than stepwise search. 

In summary we can conclude that the result of Proposition \ref{Prop1} does not always hold for generalized linear models, but that searching over a range of values of $\lambda$ and considering that model which minimizes BIC along the regularization path provides a simple and efficient strategy to overcome that problem.


\subsection{Least squares segmentation}\label{Subsec:Seg}

We finally want to discuss least squares segmentation of a one-dimensional signal, which was recently applied in the context of analyzing pathological patterns of DNA in tumor tissues \cite{rippe2012vizualization}. Let  $\boldsymbol y \in \R^n$ denote $n$ measurements which are spatially (or temporally) ordered. Then the problem of segmentation can be formalized by introducing $L_0$ penalties for changing the estimated mean between neighboring measurements,
\beq\label{Eq:segmentationLSS}
\hat{\boldsymbol \mu}= \arg \min_{\boldsymbol \mu \in \R^n} \left\{ \sum_{i=1}^{n} (y_i-\mu_i)^2 + \lambda \sum_{i=1}^{n-1} \mathbbm{1}(\mu_i \neq \mu_{i+1})\right\} \; ,
\eeq
where $\mathbbm{1}(\cdot) \in \{0,1\}$ is the indicator function. According to Remark \ref{Rem:mat} this fits into our context as a slightly generalized version of the penalized contrast (\ref{eq:pencontrast}), and like in \cite{rippe2012vizualization} we introduce the following weighted Ridge square loss as a generalization  of (\ref{eq:penARcontrast}):
\beq
\mbox{SL}(\boldsymbol \mu; \lambda,\boldsymbol w)=\sum_{i=1}^{n} (y_i-\mu_i)^2 + \lambda \sum_{i=1}^{n-1} w_i(\mu_{i+1}-\mu_i)^2 \; .
\eeq
For the corresponding AR procedure we again start with the initial weights $\boldsymbol w^{(0)} \simeq 1$ and for $k\geq 1$ perform the iterations
\beq\label{Eq:rigdeLSS}
{\boldsymbol \mu}^{(k)} = \arg \min_{\boldsymbol \mu \in \R^n} \mbox{SL}\left(\boldsymbol \mu; \lambda,\boldsymbol w^{(k-1)} \right) \; ,
\ \
w_i^{(k)}= 
\left( {\delta^2 + \left( \mu_{i+1}^{(k)}- \mu_{i}^{(k)}\right)^2}\right)^{-1} \;.
\eeq
The computations of (\ref{Eq:rigdeLSS}) can be easily solved analytically by considering the derivatives of $\mbox{SL}(\boldsymbol \mu; \lambda,\boldsymbol w)$. Minimization of the loss function then corresponds to solving the following set of linear equations:
\beq
\left\{
\begin{array}{l}
(y_1-\mu_1)+\lambda w_1 (\mu_2-\mu_1)=0 \\
(y_2-\mu_2)+\lambda w_2 (\mu_3-\mu_2)-\lambda w_1(\mu_2-\mu_1)=0 \\
\vdots \\
(y_{n}-\mu_{n})-\lambda w_{n-1} (\mu_n-\mu_{n-1})=0 \\
\end{array}
\right.
\eeq
In \cite{rippe2012vizualization} it was suggested to solve this problem using an efficient sparse LU decomposition. Here we provide a dramatically faster alternative which allows to recursively compute the solution. For $i=1,\ldots,n-1$, let us write $\mu_i=a_i+b_i \mu_{i+1}$ where $a_i,b_i \in \R$. From the linear equations above we obtain
\beq
\begin{array}{lll}
a_1=\displaystyle\frac{y_1}{1+\lambda w_1}
&
b_1=\displaystyle\frac{\lambda w_1}{1+\lambda w_1} & i=1;\\[4mm]
a_i=\displaystyle\frac{y_i+\lambda w_{i-1}a_{i-1}}{D_i}
&
b_i=\displaystyle\frac{\lambda w_{i}}{D_i}
& 1 < i < n \; ,
\end{array}
\eeq
with $D_i = 1+\lambda w_i+\lambda w_{i-1}(1-b_{i-1})$, and finally 
\beq
\begin{array}{ll}
\mu_n=\displaystyle\frac{y_n+\lambda w_{n-1}a_{n-1}}{1+\lambda w_{n-1}(1-b_{n-1})}& i=n \; , \\[4mm]
\mu_{i}=a_{i}+b_{i}\mu_{i+1} & i<n \; .
\end{array}
\eeq
Using these recursive formulas one can hence perform one update step of (\ref{Eq:rigdeLSS}) in $\mathcal{O}(n)$.
Alternatively, one can use dynamic programming to find the best solution of (\ref{Eq:segmentationLSS}) with at most $k_\text{max} \geq 1$ segments in $\mathcal{O}(k_\text{max} \times n^2)$. Such a strategy is for example explained in \citet{rigaill2010pruned} and implemented in the {\tt Segmentor3IsBack} R package \citep{cleynensegmentor3isback}.

\begin{figure}[tb!]
\begin{center}
\begin{tabular}{cc}
\includegraphics[width=0.5\textwidth]{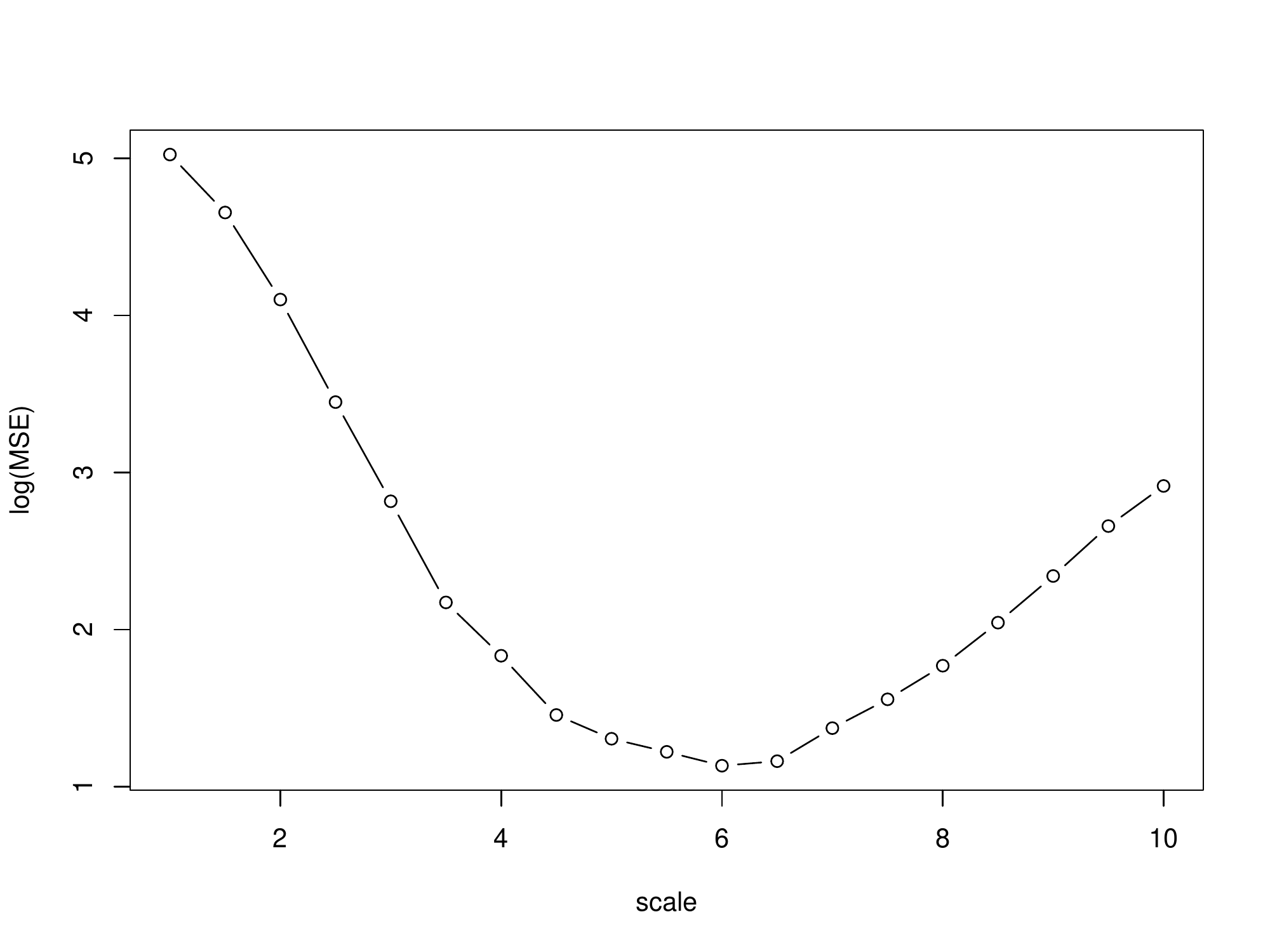}&
\includegraphics[width=0.5\textwidth]{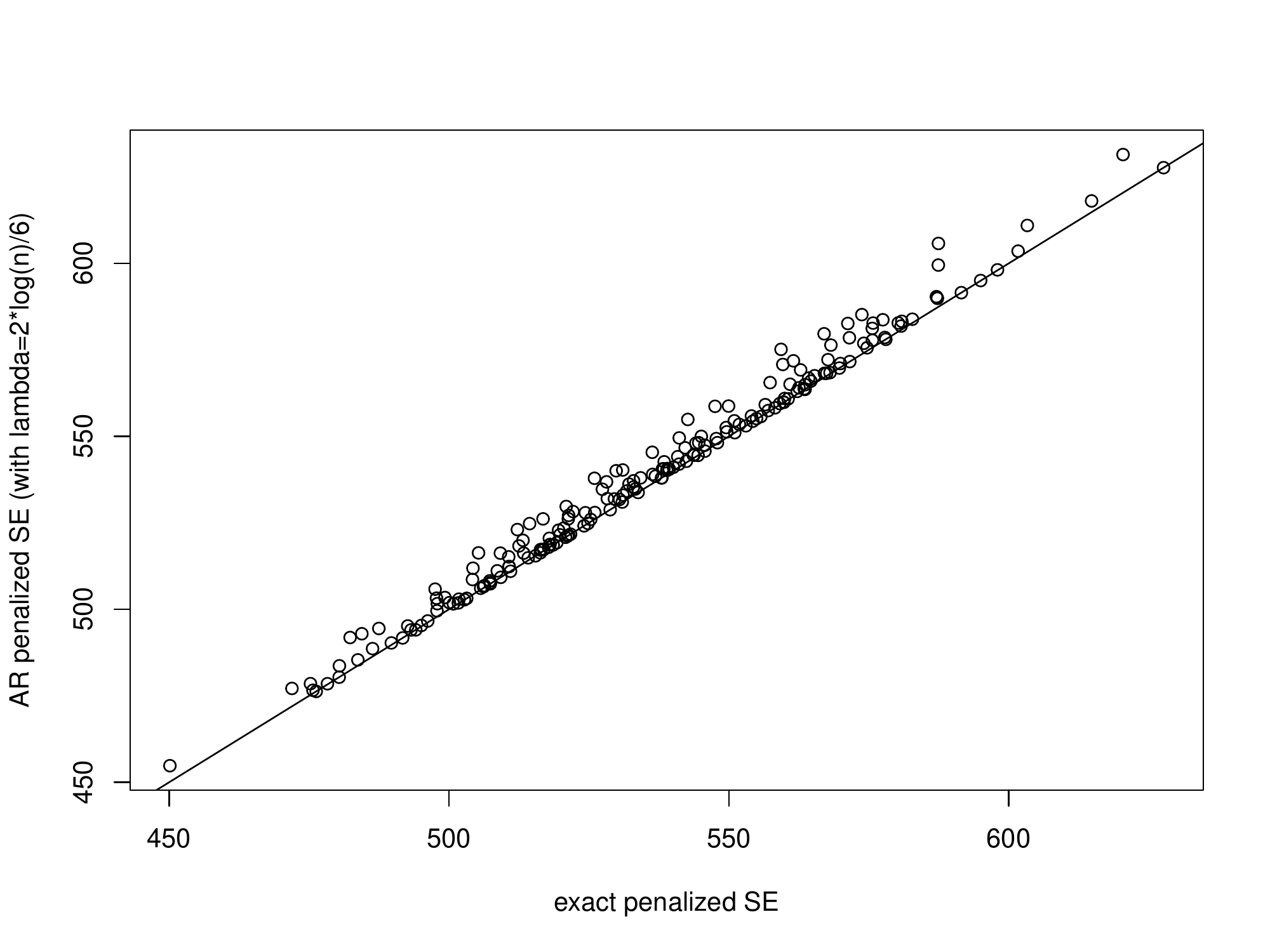}\\
(a) & (b)\\
\includegraphics[width=0.5\textwidth]{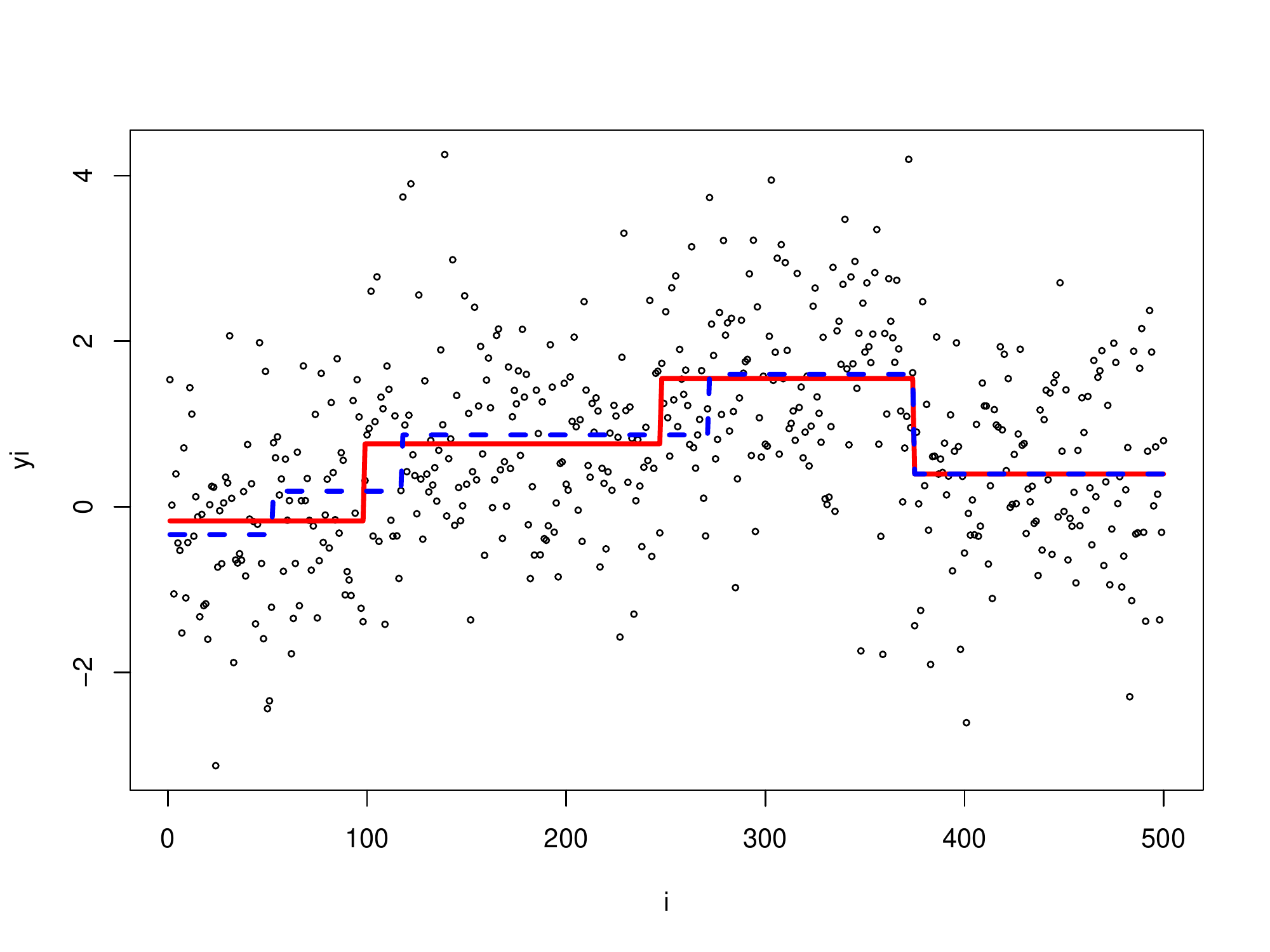}&
\includegraphics[width=0.5\textwidth]{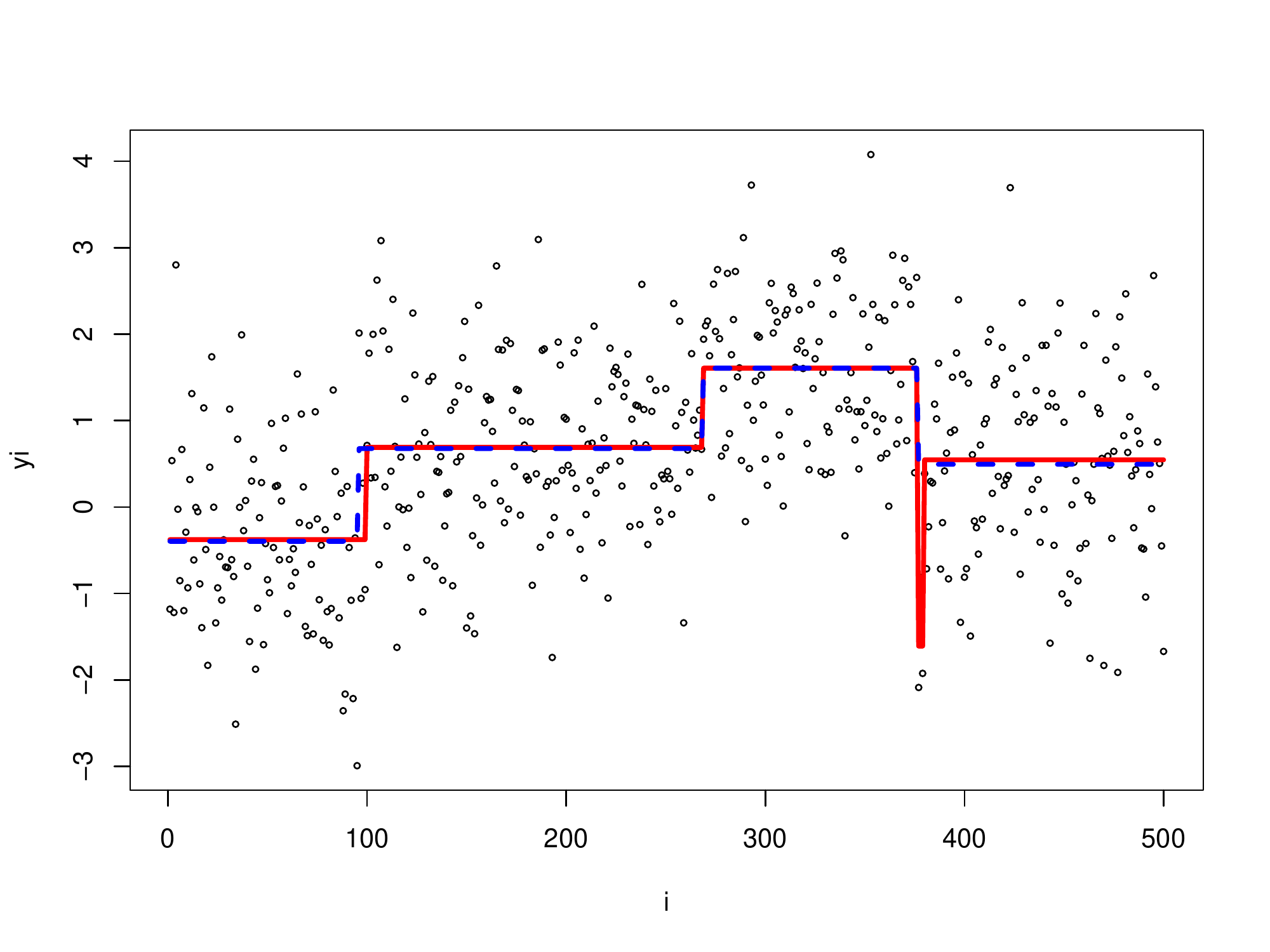}\\
(c) & (d)
\end{tabular}
\end{center}
\caption{{Comparison of exact segmentation ($\lambda=2\log n$) and adaptive Ridge segmentation ($\tilde\lambda=\lambda/\text{scale}$).
Panel (a) show the calibration of the rescaling parameter which leads to $\text{scale}=6$. Panel (b) compares the exact penalized SE to the one obtained through AR with $\lambda=2\log(n)/6$.
Panels (c) and (d) illustrate the segmentation output for two specific instances where one observes some disagreement between exact (red solid line) and  AR segmentation (blue dashed line).}}\label{fig:segmentation}
\end{figure}
In order to validate the adaptive Ridge approach in the context of least squares segmentation we will compare its performance with the exact approach in a small simulation study.
We consider a simple Gaussian design with $n=500$ consecutive measurements and three breakpoints at positions $100$, $250$ and $375$. Based on a Gaussian model 200 data sets were generated with mean values $-0.3,0.7,1.5,0.5$ in the four different segments, and a common standard deviation of $\s^2 = 1.0$.
After performing some calibration of the parameter $\tilde \lambda$ using the previously discussed warm start method of AR (Figure~\ref{fig:segmentation}a) we decided upon using the AR penalty $\tilde\lambda=\lambda/6$, where $\lambda=2\log(n)$ is the penalty of the original  criterion. This rescaling factor appeared to be quite stable for various scenarios, though perhaps increasing slightly with growing $n$ (data not shown).

We can see in Figure~\ref{fig:segmentation}b a comparison of the SE penalized criterion obtained both by exact computations and the AR method. AR clearly gives good results, although sometimes suboptimal. In Figure~\ref{fig:segmentation}c,d we give two examples of such suboptimal situations: in Figure~\ref{fig:segmentation}c, AR has two misplaced breakpoints and the selection of an additional one. In Figure~\ref{fig:segmentation}d AR missed one very small segment that was considered relevant by the exact approach. Thus although AR did not find the optimal model in terms of the criterion, its solution is in fact closer to the underlying true model.  Given the general good performance of AR one might conclude that due to its efficiency it might be preferable to looking for the exact solution particularly for large scale problems.


\section{Discussion} \label{Sec:Disc}

In this paper we have introduced the adaptive ridge procedure AR, an iterative procedure whose purpose is to solve $L_0$ penalized problems via weighted ridge optimization. The approach, recently suggested by \cite{rippe2012vizualization} in the particular context of least squares segmentation, is very similar to the iterative adaptive Lasso procedure introduced in \cite{BM08,candes2008enhancing}, with the noticeable difference that AR requires at each iteration to solve a weighted ridge problem instead of the weighted Lasso. As a result, the practical implementation of the adaptive ridge is often dramatically simpler than its adaptive lasso counterparts, and it is computationally much less expensive. This is illustrated particulary in Section~\ref{Sec:Ex}, where we provide a simple solution to the three classical problems of Poisson regression, logistic regression, and least squares segmentation.

It was pointed out in \cite{rippe2012vizualization} that the adaptive ridge approach clearly performs very well in practice, though any theoretical justifications of that behavior was missing. In this paper we partially addressed this problem by studying the dynamics of AR in the particular case of orthogonal linear regression (with known variance). In this context we derived explicit conditions for the convergence of AR and proved that the adaptive ridge penalty needs to be four times smaller than the original $L_0$ penalty to give the same results. According to our simulations this scaling factor of $1/4$  worked quite well also in case of non-orthogonal linear regression,  as long as the correlation between covariates was not too high. In case of highly correlated regressors, as well as for $p \gg n$, further investigation might be necessary, but in general such rescaling offers a natural way to select adaptive ridge penalties by targeting classical $L_0$ penalty schemes like AIC and BIC, or in a high-dimensional setting the more recently suggested mBIC.  

Furthermore the AR procedure, just like the lasso, allows to take advantage of warm starts to compute efficiently the entire solution surface for a sequence of penalties. This gives the possibility to select the most appropriate penalty of AR without any need to know the rescaling scheme. Note that for the adaptive ridge we have to consider increasing penalty values, whereas for the lasso one usually considers decreasing penalty values.

In summary the AR procedure suggested in this paper is quite straightforward to understand and implement, can be easily combined with iterative optimization procedures like Newton-Raphson, and offers efficient ways to compute entire solution surfaces. We hope that this paper could be a first step to learn more about the theoretical properties of this method, which definitely seems to be worth of further investigation.

\bibliographystyle{chicago}
\bibliography{ARR_FF}

\end{document}